\documentclass[12pt,preprint]{aastex}
\usepackage{graphicx,amssymb,amsmath,times}
\usepackage{natbib}
\usepackage{longtable,threeparttablex,booktabs}
\usepackage{footnote}
\usepackage{caption}
\usepackage{lscape}
\usepackage[singlelinecheck=off]{caption}
\usepackage{relsize}
\usepackage{amsmath, amssymb} 
\newcommand{\lsim}{{\lower.5ex\hbox{$\; \buildrel < \over \sim \;$}}}
\newcommand{\gsim}{{\lower.5ex\hbox{$\; \buildrel > \over \sim \;$}}} 
\usepackage{url}
\usepackage[T1]{fontenc}
\usepackage[latin9]{inputenc}
\usepackage{multirow}
\usepackage{rotating}
\usepackage{geometry}
\usepackage{changepage}
\usepackage{txfonts}
\usepackage{rotating}
\def\ltsima{$\; \buildrel < \over \sim \;$}
\def\simlt{\lower.5ex\hbox{\ltsima}}
\def\gtsima{$\; \buildrel > \over \sim \;$}
\def\simgt{\lower.5ex\hbox{\gtsima}}

\def\farcm{\hbox{$\mkern-4mu^\prime$}}
\def\farcs{\hbox{$^{\prime\prime}$}~}

\shorttitle{Herschel observations of HII region NGC~3603}
\shortauthors{Di Cecco et al.}

\begin{document}
\title{Herschel Far IR observations of the giant HII region NGC~3603\thanks{{\it Herschel} is an ESA space observatory with science instruments provided by European-led Principal Investigator consortia and with important participation from NASA.}}

\author{Alessandra Di Cecco\altaffilmark{1,2,3}, 
Fabiana Faustini\altaffilmark{2,3}, 
Francesco Paresce\altaffilmark{4}, 
Matteo Correnti \altaffilmark{5} and 
Luca Calzoletti\altaffilmark{2,3,6}}
   
\altaffiltext{1}{INAF - Osservatorio Astronomico di Teramo, Via  Mentore Maggini snc, I-64100, Teramo, Italy, dicecco@oa-teramo.inaf.it}

\altaffiltext{2}{ASDC - ASI Science Data Center, Via G. Galilei snc, I-00044 Frascati (RM), Italy}   

\altaffiltext{3}{INAF - Osservatorio Astronomico di Roma, Via  di Frascati 33, I-00040 Monte Porzio Catone (RM), Italy} 

\altaffiltext{4}{INAF - Istituto di Astrofisica Spaziale e Fisica Cosmica, Via Piero Gobetti 101, I-40129  Bologna, Italy}

\altaffiltext{5}{Space Telescope Science Institute, 3700 San Martin Drive, Baltimore, MD
21218, USA}

\altaffiltext{6}{European Space Astronomy Centre, Herschel Science Centre, Villanueva  de la Canada, E-28691 Madrid, Spain}

\date{\centering drafted \today\ / Received / Accepted }

\begin{abstract}
   
We observed the giant HII region around the NGC~3603~YC with the 5 broad bands ($70, 160, 250, 350, 500~\mu$m) of the SPIRE and PACS instruments, on-board the Herschel Space Observatory.   Together with what is currently known of the stellar, atomic, molecular and warm dust components, this additional and crucial information should allow us to better understand the details of the star formation history in this region. The main objective of the investigation is to study, at high spatial resolution, the distribution and main physical characteristics of the cold dust. By reconstructing the temperature and density maps, we found  respectively a mean value of 36~K and  $log_{10}N_{H}=22.0\pm0.1$~cm$^{-2}$. We carried out a photometric analysis detecting  107 point-like sources, mostly confined to the North and South of the cluster. By comparing our data with SED models we found that 35 sources are well represented by YSOs in early evolutionary phases, from Classes~0 to Class~I. The Herschel detections also provided far-IR counterparts for 4 H2O masers and 11 objects previously known from mid-IR observations. The existence of so many embedded sources confirms the hypothesis of an intense and ongoing star formation activity in the region around NGC~3603~YC. 

\end{abstract}

\maketitle

\section{Introduction}
\label{Intro}

The giant Galactic HII region NGC~3603 is located at a distance of 7 $\pm$1 Kpc and is part of the RCW 57 complex South-East (SE) of Eta Car, in the Carina arm. The young, bright, compact stellar cluster (NGC~3603~YC, l=291.62 $^{\circ}$, b=-0$^{\circ}$.52) lying  roughly at the center of this region and has a bolometric luminosity $\sim$100 times that of the Orion Cluster. With a total mass density in the core of $\sim6~10^4 M_{\Sun}/pc^3$ \citep{harayama08}, NGC~3603~YC  is usually placed in the category of super star clusters of the type more commonly seen in starburst regions of young galaxies. A recent study by \citet[][hereinafter F14]{fukui} suggests that this cluster likely originated by a cloud-cloud collision that occurred $\sim 1$~Myr ago. While there is a considerable amount of information available about its stellar population, that has been used to study its structure and evolution \citep{correnti12,beccari10,rochau}, detailed  knowledge about the massive \citep[$\sim4~10^5 M_{\Sun}$,][]{grab88} cloud complex surrounding it is still rather limited. Studies of its atomic and molecular gas and dust properties are rare, and only small areas at or near the center have been observed at sufficient  angular resolution in the near and mid-IR \citep[][hereinafter NS03]{nurn02,nurn03}. Recent larger scale IR surveys  include: the IRAS - HiRes observations at 12 to 100 microns of a region of $\sim 20~\farcm$x~$20~\farcm$ centered on the  cluster, with a spatial resolution of approx 2~\farcm ~($\sim$~4 pc);  a radio survey using the CS (2-1) and (3-2) lines at 98 and 147 GHz, over an area oriented North-South (NS) of $5.8~\farcm $x$16.7~\farcm$ ($\sim$~12~pc~x~34~pc) and with a resolution of $\sim35\farcs$ (1pc) \citep[][hereinafter N02]{nurn02b}; an MSX satellite mid IR survey of the entire cloud region ($2^{\circ}$x$2^{\circ}$) in 4 wavelength bands at very low resolution between  8 and 20 microns \citep{WangChen10}. These measurements found dense molecular cores, embedded near and mid IR sources as well as several water, methanol and OH maser sources \citep[][hereinafter B10]{caswell04,breen10}, clearly implying that star formation is still ongoing in this region. They also found large amounts of diffuse emission closely correlated with ionized material and identified several shocks and ionization fronts resulting from the impact of fast stellar winds, as well as, ionizing photons on the adjacent gas and dust originating from the massive cluster stars. \\
\noindent
The complexity of this diffuse component (pillars, shocks, ionization fronts) and the large diversity of compact sources makes a full understanding of the star formation history, physical properties and future evolution of NGC~3603 very difficult to unravel, especially since no sub arc-second resolution observations at far IR wavelengths of the entire complex were available until recently. To fill this crucial gap, we present in this paper the results of the first observations of NGC~3603  performed with the ESA Herschel Space Observatory \citep{pilbratt10}, employing Herschel's large telescope and powerful science payload to do imaging and photometry using the PACS \citep{pog10} and SPIRE  \citep{griffin10} instruments. Using all five photometric bands, we can detect and characterize different types of point sources, including evolved early type stars, relatively young late type stars, candidate protostars and maser sources. Moreover, we are able to determine colour temperature and column density maps in the central regions of the complex. 
The plan of the paper is the following: in Sect.\ref{obs2} we  show the Herschel images and we study the diffuse dust component by reconstructing the maps of temperature and column density, in Sect.\ref{pointsed} we present the photometry and the classification of the point-like sources, in Sect.\ref{maser1} we compare our detections with previous studies and, finally, in Sect.\ref{conclusion} we present a summary and conclusions.

\section{Observations and data reduction}
\label{obs}

\subsection{Analysis of the cloud}
\label{obs2}
\subsubsection{Spatial distribution}
\label{obs3}

The cloud complex surrounding NGC~3603~YC  was observed by Herschel during the Hi-GAL Survey \citep[][hereinafter MO10]{mol10, mol10b}, a Key Programme that mapped the Galactic Plane in 5 photometric bands (70, 160, 250, 350 and 500 $\mu$m). The images were acquired by using the SPIRE and PACS photometers in the SPIRE/PACS Parallel Observing mode.  NGC~3603 complex lies on the overlapping region between two consecutive tiles of the Hi-GAL Survey, i.e. the tile centered at b=0$^{\circ}$ and l=290$^{\circ}$ and the tile centered at b=0$^{\circ}$ and l=292$^{\circ}$ (a detailed description of the observing strategy is given in MO10).
In order to obtain a complete coverage of the NGC~3603 region and to provide higher quality maps, we combined both tiles by using the Unimap mapmaker \citep{piaz12}, starting from the Level 1 products previously obtained by the UniHIPE tool\footnote{http://herschel.asdc.asi.it/index.php?page=unimap.html}. 
The pixel scales used for the production of the maps are those suggested by \citet{traf11} and the astrometric accuracy is 2\farcs for each band. 
The FWHM (Full-Width-Half-Maximum) approaches the diffraction limit and is equal to the instrumental beam (see Tab.\ref{tab1}). For each band the sensitivity limit is given by the instrumental confusion noise. Final flux of the maps was obtained by using the Hi-Gal calibration tiles (MO10).

\begin{figure*}[t]
\includegraphics[width=16cm]{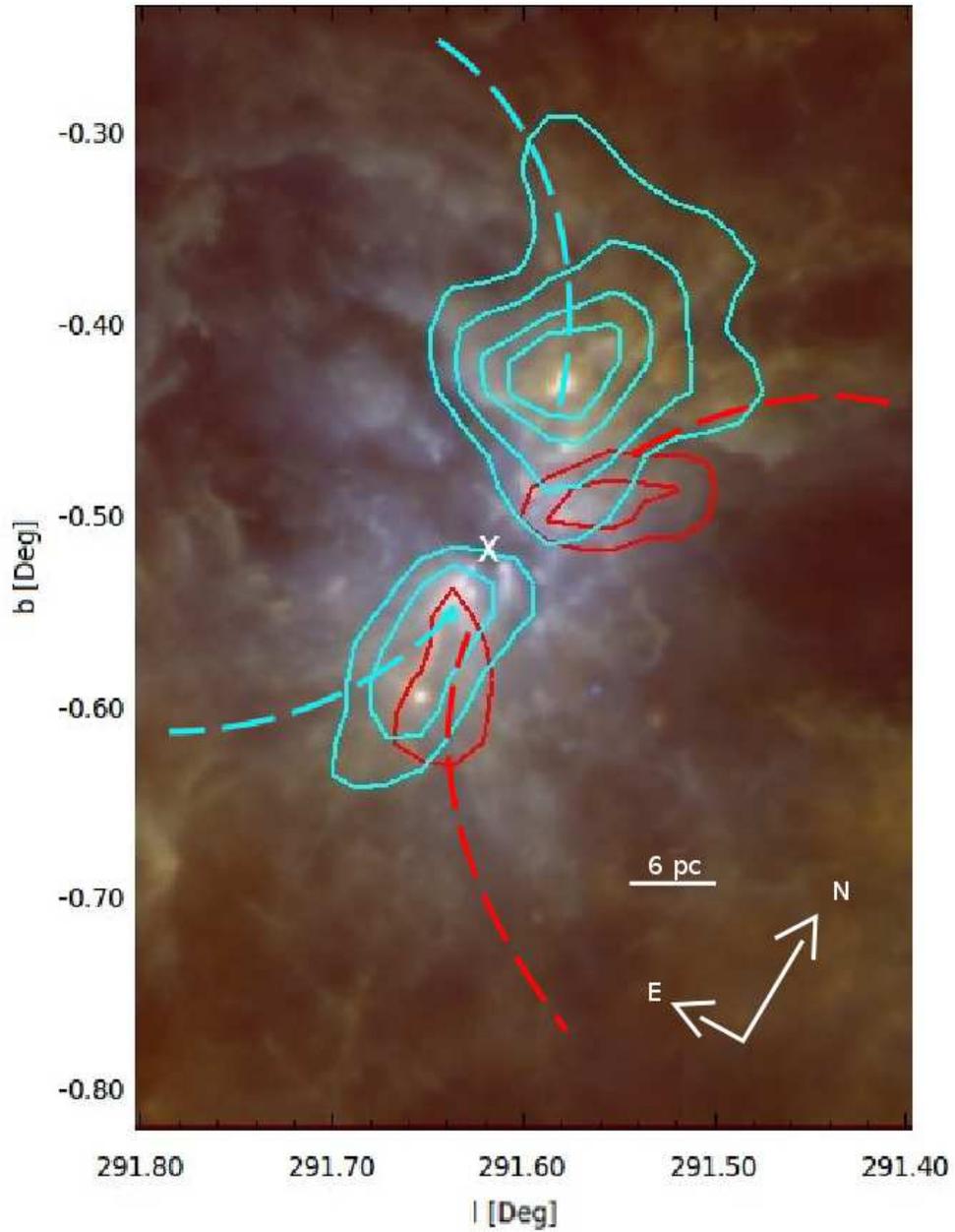}
\caption{RGB image obtained by combining the 70, 160 and 250 $\mu$m Herschel bands. The white cross marks the position of the NGC~3603~YC. The blue ($15, 42, 85, 125$ K~km/s) and red ($6, 8$  K~km/s) contours account for the F14 blue-shifted and red-shifted clouds.  The red and blue dashed lines outline the lobes of the proposed hourglass-shaped morphology. }
\label{f1}
\end{figure*}

We selected a sky area of $\sim$ 24~\farcm~x~36~\farcm around NGC~3603~YC in order to exclude the contamination by the nearby cluster NGC~3576 \citep{persi94, georgelin, town11}.
By combining the 70, 160 and 250 $\mu$m bands, we present the Red-Green-Blue (RGB) image in Fig.\ref{f1}. 
 This image reveals bright compact high-density structures located towards
the North and South of the cluster.
Among these structures the MM1 and MM2 pillars detected by N02, also observed in optical bands by \citet{brandner00}, are clearly visible. 
These compact regions are supposed to be sites of ongoing star formation activity \citep[N02, NS03 and ][hereinafter P11]{xiao11}, likely triggered by the strong radiation field of NGC~3603~YC.
 In addition, the RGB image suggests the presence of two lobes
with an hourglass-shaped morphology oriented in East-West (EW) direction,
which are mostly void of dust.
The remnant dust inside these lobes appears hotter
(bluer colours in the RGB image) and more diffuse in comparison to the
colder dust (redder colours) associated with the compact structures in
NS direction. 
To explain (the potential origin of) the hourglass-shaped morphology,
we refer to the recent work of F14, who suggested a cloud-cloud collision for the origin of NGC~3603~YC. 
In particular, in Fig.\ref{f1} we over-plotted the contours of the two parent clouds identified by F14 (see their Fig.9, Tab.2 and Tab.4).
 We found that these parent clouds not only overlap with the NS oriented high-density structures, but also seem to confine the EW oriented lobes and to outline the hourglass morphology. 

\begin{figure*}
\centering
\includegraphics[width=14cm,height=10cm]{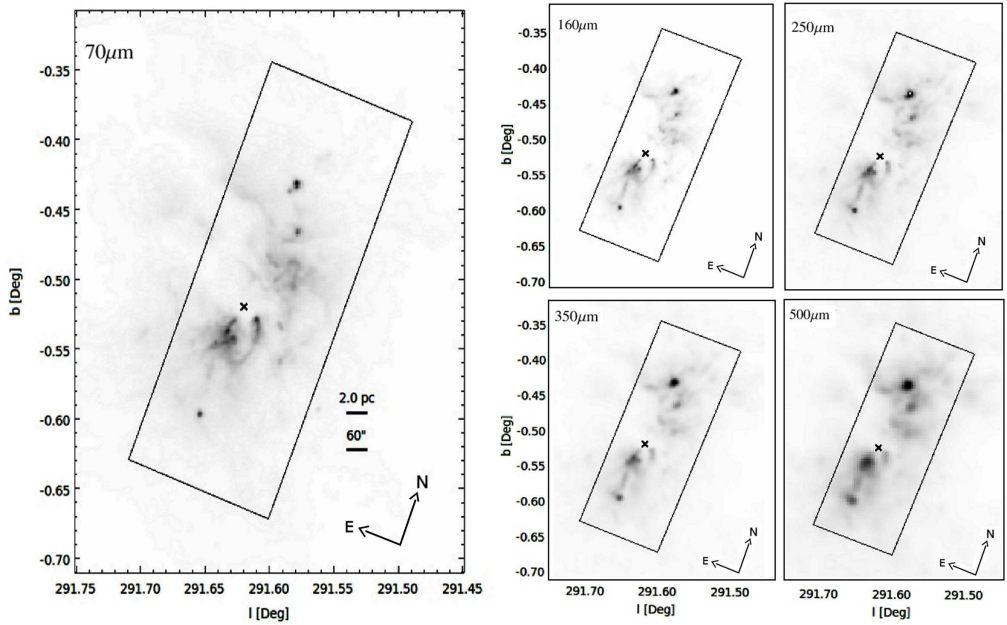}
\caption{Herschel 70, 160, 250, 350 and 500 $\mu$m images for NGC~3603. The N02 region lies within the box. }
\label{f2}
\end{figure*}

To exclude the foreground/background FIR regions, in the next sections we focused on the sky area studied by the spectroscopic observations of N02, also included in the area observed by F14.
We present the Herschel single-band images and the N02 validity region in Fig.\ref{f2}.
 The observational details of the Herschel images such as pixel scales, FWHMs and sensitivity limits are listed in Tab.\ref{tab1}. In the table we also anticipate the details of the MSX \citep[Midcourse Space Experiment;][]{egan03} and WISE \citep[Wide-field Infrared Survey Explorer;][]{wise} images used in Sect.2.2. 
The MSX images were taken at (the wavelengths) 8.3, 12.1, 14.7 and 21.3 $\mu$m using the scan observing mode, while the WISE images were acquired in freeze-frame scan mode at 3.4, 4.6, 12 and 22 $\mu$m.  However, the data taken in the last two WISE bands (12 and 22 $\mu$m) are completely saturated, and we do not use them for our analysis. The astrometric accuracy of MSX is of the order of $\sim$ 2\farcs --5\farcs, while for WISE it is $\sim$ 0.2\farcs.  The MSX and WISE calibrated images were retrieved from the NASA/IPAC Infrared Science Archive\footnote{http://irsa.ipac.caltech.edu}. 

\subsubsection{Temperature and density maps}

\begin{figure*}[t]
\centering
\includegraphics[scale=0.3]{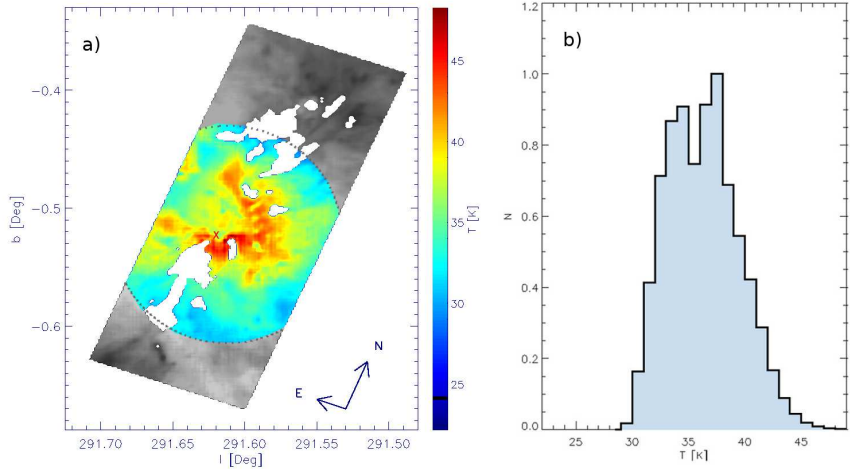}
\caption{Panel a) shows the temperature map based on 70-160 $\mu$m flux ratio. The coloured area is the validity region. White regions mask the high density structures  (see text for details). -- Panel b) shows the histogram of the temperature map.}
\label{f3}
\end{figure*}

In order to study the physical and chemical properties of the dust in the N02 region, we reconstructed the maps of the temperature and hydrogen column density.
We calculated the local temperature ($T$) by using the following flux ratio formula:  
\begin{equation}
\frac{I_{\lambda_{1}}}{I_{\lambda_{2}}}\propto(\frac{\lambda_{2}}{\lambda_{1}})^{\beta}\frac{exp(\lambda_{T}/\lambda_{2})-1}{exp(\lambda_{T}/\lambda_{1})-1},
\label{eq:1}
\end{equation} 
where $I_{\lambda}$ is the flux density measured at wavelength $\lambda$, and $\lambda_{T}=hc/kT$, where $hc/k$ are the constants of Planck's radiation formula.
In this equation, the temperature and the spectral index $\beta$ are supposed to be constant along the line of sight, and for a common composition of silicate and graphite dust, we assumed $\beta=2$ \citep{ward02, draine84}.
Due to the higher resolution at the shorter wavelengths, in Eq.\eqref{eq:1} we used the 70  and 160 $\mu$m flux ratio. To compare the intensity of different bands, we generated a $70~\mu$m image convolved with a $160~\mu$m kernel as suggested by \citet{aniano11}. 
We obtained the final temperature map by comparing the flux ratio with the values of a theoretical grid, but before accepting the results, two issues need to be considered. The first issue is the potential impact of the IR emissions by interstellar dust and the second is the validity range of the above equation with different optical depths. \\
Concerning the IR emissions, while the polycyclic aromatic hydrocarbons (PAH) show lines around $6-17~\mu$m and thence do not affect the Herschel bands, a contribution in the $70~\mu$m band could come from the very small grains (VSG)  that do not achieve an equilibrium temperature \citep[][hereinafter DL07]{draine07}. 
However, in conditions of a strong interstellar radiation field (ISRF) as in several high-mass star forming regions, the VSG approach a steady state temperature (DL07).
By using Spitzer Infrared Spectrograph (IRS) data, \citet[][hereinafter L07]{lebo07,lebo08} studied the hardness of the NGC~3603 ISRF by investigating the strength of the emissions of PAH and several forbidden lines. In particular, L07 used the line ratio [SIV/NeII] as tracer of the ISRF and they found that its hardness decreases with the increasing cluster radial distance. The authors also found that the PAH/VSG ratio  anticorrelates with the ISRF, by suggesting that the VSG emission lies in the central regions where the PAH are destroyed by high energy photons.
The NGC~3603 ISRF is dominated by the FUV flux generated by the massive O and B type stars \citep{melena08} and  quantitatively, \citet[][hereinafter R11]{roll11} calculated the flux assuming a value of $\chi=9.4~10^4~d^{-2}$  (with $d$ in parsec) in units of a Draine field \citep{draine78}.
In order to impose spatial limits for the temperature map we chose an accuracy better than $5\%$, a value that, as discussed by \citet{preibisch12}, by considering a variation of the VSG of a factor of 10, is achieved when the IRSF is 100 times the value of that of the solar neighborhood. To this purpose, we limit the area at a radial distance of $\sim 5$~\farcm~ from  NGC~3603~YC. \\
The second issue concerns the validity of the Eq.1 only in the optically thin regime. Considering the $70~\mu$m optical depth, we assumed that the temperature values are not correct in the optically thick regions characterized by $log_{10}N_H>22.7$~cm$^{-2}$ (see the following discussion on the column density). \\ 
Taking into account these considerations we masked the invalid areas and we show the resulting temperature map in panel a) of Fig.\ref{f3}. The validity area is covered by $\sim$10,000 pixels.
We found that the temperature range is 30~K~$\le T \le$ 47~K with the lowest values in the outermost regions of the cloud. We also noted that the temperatures are not symmetrically distributed around the NGC~3603~YC and the main part of the hotter ($\sim40-45K$) structures are grouped towards the West. 
Note that this result is not affected by the beam size because the Herschel beam at 160~$\mu$m is of the order of 13\farcs, significantly smaller than the observed cloud. 
In order to investigate the temperature distribution, we reconstructed the histogram of the map in the panel b) of Fig.\ref{f3}. We found that the histogram mode and median are similar and equal to 36$\pm1$~K, in good agreement with the results (35--37 K) provided by R11 and \citet{WangChen10}.

\begin{figure*}[b]
\centering
\includegraphics[scale=0.3]{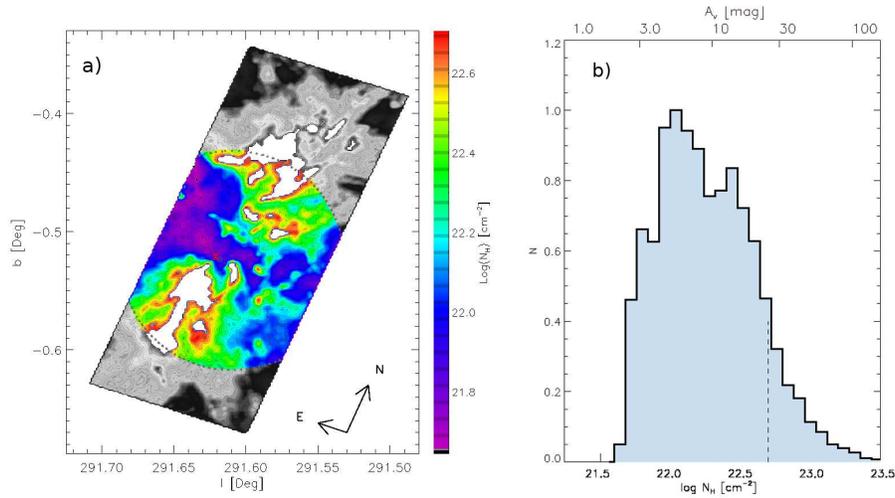} 
\caption{Panel a) shows the column density map reconstructed by using the temperature values of Fig.\ref{f3}.  -- Panel b) shows the histogram of the density map.}
\label{f4}
\end{figure*}

By using the temperature map values, we derived the hydrogen column density (N$_H$) with the following formula:
\begin{equation}
N_H=2~N_{H_2} = 2~\frac {I_{\nu}~R}{\Omega\mu_{H_2}~m_H~k_{\nu}~B_{\nu}(T)}
\end{equation}

where $I_{\nu}$ is the intensity at $160~\mu m$, $R$ is the gas-to-dust ratio (we assumed $R=100$), $\Omega$ is the beam solid angle, $\mu_{H_2}$ is the mean molecular weight per hydrogen molecule ($\mu_{H_2}=2.8$), $m_H$ is the hydrogen atomic mass ($m_H=1.66~10^{-24}$g), $B_{\nu}(T)$ is Planck's equation at temperature $T$, and $k_{\nu}$ is 
the dust opacity. We assumed $k_{\nu}=40.9$ cm$^2$/g, as suggested by \citet[][hereinafter OH94]{oh94} for protostellar cores formed by thin ice mantle grains with density $n=10^6$ cm$^{-3}$.
Taking into account that the uncertainty on the Herschel intensity at $160~\mu$m is low ($\sim$ few percent), in Eq.2 the total uncertainty is dominated by the dust opacity whose error is of a factor of two.
The optically thin regime assumed for the temperature map was established by using the value of $N_H$ for which the $70~\mu$m optical depth $\tau \le0.1$. To be conservative,
we assumed an uncertainty of a factor of two on the $k_{70\mu m}$ opacity (OH94) and we obtained that the optically thin regime is defined by $log_{10}N_H\leq22.7$~cm$^{-2}$. We present the column density map in panel a) of Fig.\ref{f4}. We found  that the higher density regions are clumped North and South of NGC~3603~YC. In panel b) of Fig.\ref{f4} we show the histogram of the column density map, whose mode is $log_{10}N_{H}=22.0\pm0.1$~cm$^{-2}$  and median is $log_{10}N_{H}=22.2\pm0.1$~cm$^{-2}$. These results well agree with the value of $log_{10}N_{H}=22.0\pm0.1$~cm$^{-2}$ provided by P11 with the $H_\alpha/P\beta$ flux ratio, obtained with optical and near-IR HST/WFC3 images. 
Taking into account the relation between the column density and the visual extinction \citep[$A_v \sim N_H/(1.9~10^{21})$,][]{bohlin78}, the mode of the column density histogram suggests a value of $A_v=5.3$ mag in agreement with the literature values \citep[$4.5<A_v<5.7$ mag,][]{brandl99, moffat83, melnick89, correnti12}. By plotting in the histogram the pixels in the optically thick regime ($log_{10}N_{H}>22.7$~cm$^{-2}$), we evaluated their number of the order of $\sim9\%$ of the entire useful region. 

In order to check the results obtained under the assumptions adopted for the 70$\mu$m data, we reconstructed the temperature map by using the other four (160, 250, 350, 500 $\mu$m) Herschel bands (never affected by VSG). We convolved and regridded the 160, 250, 350 $\mu$m images with the kernel and pixel scale of the $500~\mu$m, obtaining four images with $\sim3,500$ pixels in the N02 region. For each pixel we ran the IDL MPFITFUN routine based on the least-square fitting between the observed and expected values of a fixed function \citep[see Sect.3.1.2 in][]{etxaluze}. To model the dust continuum emission we used the Spectral Energy Distribution (SED) described by the modified black body:

\begin{equation}
I_{\nu}=\Omega B_{\nu}{(T)} (1-e^{-\tau_{\nu}})~~ \textrm{, where}~~ \tau_{\nu}=\tau_{0}(\frac{\nu}{\nu_0})^\beta
\end{equation}

\begin{figure*}
\centering
\includegraphics[width=14cm]{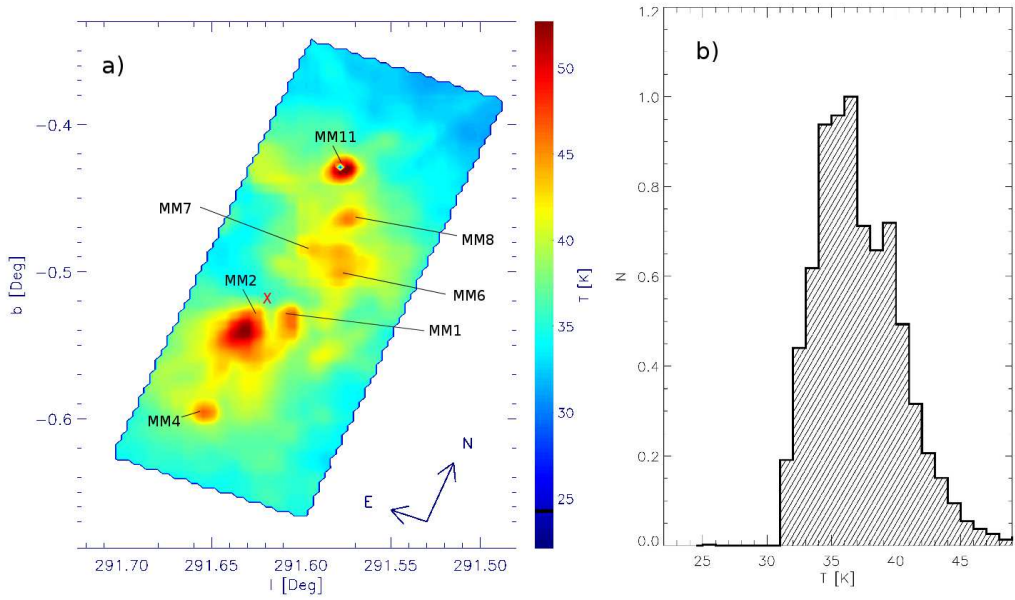}
\caption{Panel a) shows the temperature map reconstructed by using the SED fitting method with the 160, 250, 350 and 500 $\mu$m images. Several clumps from N02 are marked. - Panel b) shows the histogram of the temperature map.}
\label{f5}
\end{figure*}

To be consistent with the Eq.1 we fixed the dust spectral index at $\beta = 2$. The outputs are the optical depth $\tau_{0}$ (at the chosen reference wavelength of $250~\mu$m) and the temperature.
Note that in contrast to the Eq.1, the Eq.3 is valid also in the optically thick regime. We show the temperature map in panel a) of Fig.\ref{f5}. We find that the temperature range is 30~K~$\le T \le$~53~K, in agreement with the results (30-50K) provided by F14 investigating the intensity ratio of CO lines. 
As shown by the temperatures reconstructed with the $70-160~\mu$m intensity ratio, the map in Fig.\ref{f5} confirms that the dust between the range 40~K~$\le T \le$~45~K is mainly located  Westward of the cluster.
Last map shows that the temperatures follow the same spatial distribution of the CS(2-1) emissions detected by N02 (see their left panel in Fig.3), and likely associated to star forming regions. In particular, the hottest regions are clearly associated with the highest molecular emissions, as marked by the clumps shown in Fig.\ref{f5}. This evidence confirms that the dust around these clumps is probably heated by the dynamical and physical processes which take place during the star formation.\\
In the MM1 and MM2 pillars we found a temperature of 44~K and 47~K respectively, in agreement with the values provided by R11 ($T_{MM1}=43$~K, $T_{MM2}=47$~K). Taking into account the $\tau_{0}$ values of the pillars, we calculated the half column density $log_{10}N_{H_2}^{MM1}=22.8$~cm$^{-2}$ and $log_{10}N_{H_2}^{MM2}=23.2$~cm$^{-2}$.  
These results are consistent with the literature values provided by P11 and N02 ($log_{10}N_{H_2}\sim22.6-23$~cm$^{-2}$), as well as with the density values found for the star formation activity at the edges of HII regions \citep{urq09}. The strong radiation front generated by NGC~3603~YC is likely ionizing the MM1 and MM2  pillar heads which in turn are the sites of strong star formation processes (P11, NS03), as also supported by the methanol and water maser emissions \citep{caswell89,depree99}. \\
The histogram of the temperature map is shown in panel b) of Fig.\ref{f5}. The mode and the median are similar and equal to $36\pm1$~K, confirming the value obtained by using the $70-160~\mu$m intensity ratio and the literature results \citep[R11,][]{WangChen10}. Owing to the larger area covered by the temperature map in panel a), the histogram also suggests that the mean temperature of the cloud is almost homogeneous across the N02 region.

\subsection{Point-like sources and spectral energy distributions}
\label{pointsed}
 In the N02 region, we carried out an accurate photometric analysis of the Herschel data by using the CuTEx algorithm \citep[][hereinafter MO11]{mol11}, both for the source detection and flux extraction.
We imposed a detection limit of $3\sigma$ above the background in the 2$^{nd}$-order CuTEx derivatives images, and estimates the flux uncertainty of the order of 10\% (see MO11). Owing to the decrease of the resolving power with the increasing wavelength, several clumped sources clearly resolved at the shorter wavelengths remain unresolved at the longer wavelengths ({\it multiple association}); for the same reason the achieved accuracy of the sources position is wavelength dependent too (see the FWHMs in Tab.1). The CuTEx algorithm was also applied on the complementary MSX images (8.3, 12.1, 14.7 and 21.3 $\mu$m) to identify mid-IR counterparts to the detected Herschel point-like sources (within 1 FWHM).
For the final catalog, we rejected all sources detected in only one Herschel band without MSX counterpart; hence we ended up with 107 point-like sources, as listed in Tab.2. 
In the following, we refer to these sources with the Her-ID number.
The coordinates are based on the shortest-wavelength Herschel detection.

In order to reconstruct the SEDs, as described following, we used the WISE images (3.4 and 4.6 $\mu$m) to find near-IR counterpart for the sources in common between Herschel and MSX. We provided the WISE photometry only for these common sources for a twofold reason: 1) the MSX mid-IR bands lie between the Herschel and the WISE bands, providing a better association between the Herschel far-IR and the WISE near-IR detections; 2) the SED models are very sensitive to the MSX data and so the MSX fluxes impose essential constraints in discriminating between the numerous models. As listed in Tab.2, we found 18 WISE counterparts.

Note that owing to the lower resolution and sensitivity of MSX respect to the Herschel (70 and 160 $\mu$m)  and WISE, 
several sources could be masked by the MSX confusion noise. This fact limits the detection of sources  especially in the longest wavelength (21.3 $\mu$m), where the sensitivity limit increases by a factor of $\sim$~6 when compared to the shortest  wavelength (8.3 $\mu$m). 
To overcome this problem, for several sources, we imposed constraints in the mid-IR by calculating the upper limit of the flux (F$^U$) in the MSX bands. 
In particular, we used the following formula: 

\begin{equation}
F^U = \frac{3~rms~FWHM~\sqrt{2\pi}}{2.36}
\end{equation}

where {\it rms} is the root-mean-square flux noise calculated inside a circle centered at the (Herschel) source positions, with radius equal to the FWHM of the MSX data.  

\begin{landscape}
\begin{figure*}
\centering
\includegraphics[width=1.\columnwidth]{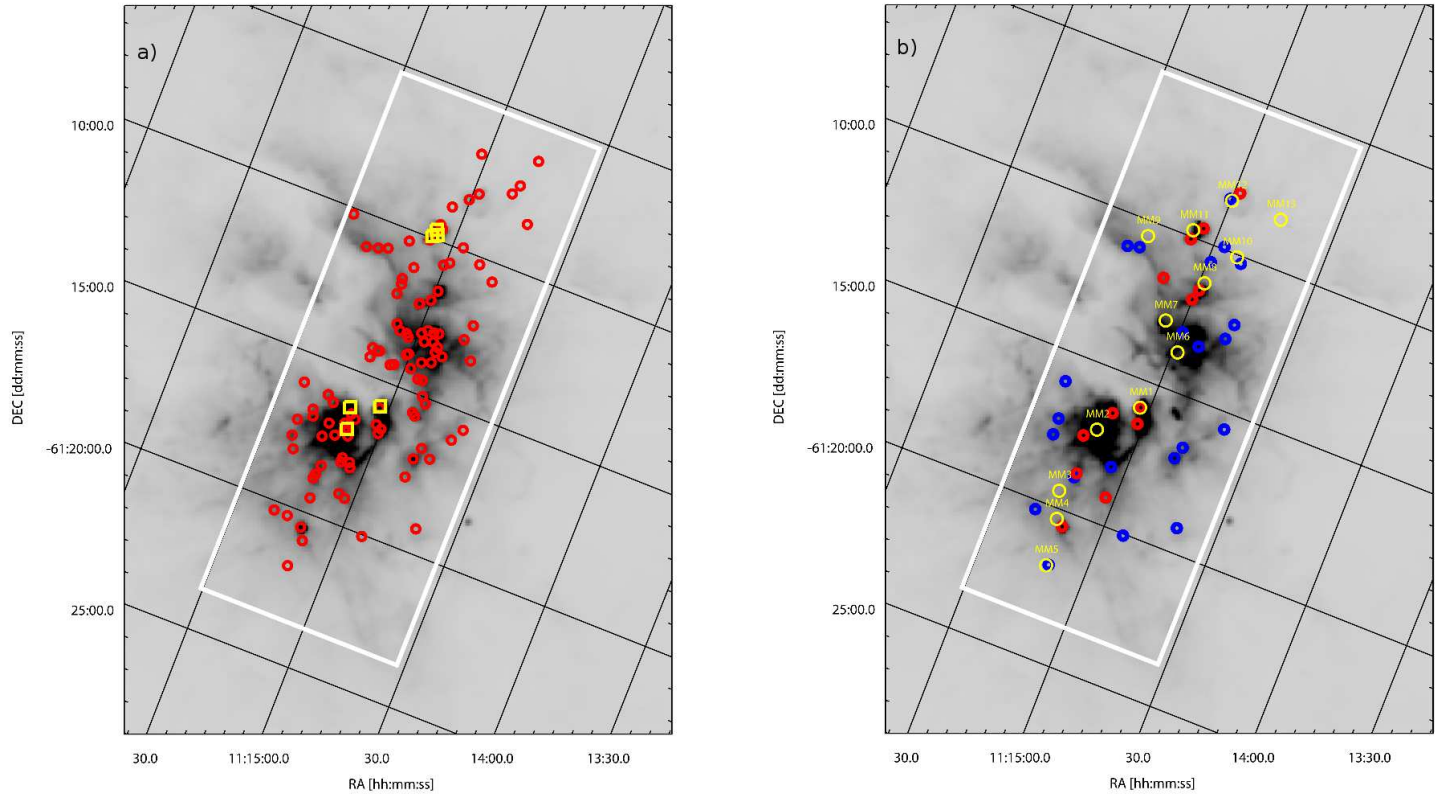}
\caption{Panel a) shows the 107 detected sources (red circle), while the (yellow) boxes mark the B10 masers. -- Panel b) shows the 35 YSO fitted sources: the red circles mark the higher mass objects ($M> 16M_{\Sun}$) and the blue circles mark the lower masses ($M<16M_{\Sun}$). The N02 clumps are also shown (yellow circles).}
\label{f6}
\end{figure*}
\end{landscape}

We plotted the positions of the detected sources on the 70~$\mu$m image in panel a) of Fig.\ref{f6}. In the same figure, the yellow boxes mark the position of the masers detected by B10 that we will discuss in detail in Sect.\ref{maser1}.
With respect to the NGC~3603~YC the Herschel detections mostly lie along the NS direction, near the positions of the molecular clumps detected by N02. In particular, we find over-densities of sources in proximity of the cluster, close to MM1, MM2, MM6 and MM7 clumps.
 The presence of these compact sources supports the hypothesis
of ongoing star formation in these regions (N02).\\

In order to investigate the evolutionary phase of the detected sources, we compared their observed SED with the models of young stellar objects (YSO) provided by the online fitting tool of \citet[][Robitaille et al. 2007, hereinafter RB07]{rob06}.
The theoretical grid covers a wide range of stellar masses (from 0.1 to 50~M$_{\sun}$) and is based on pre-computed two-dimensional (2D) radiation transfer models \citep{whi03,whi03b} with a large set of parameters. The parameters fall into three categories related to the central source (stellar mass, radius, and temperature), to the infalling envelope (envelope accretion rate, outer radius, inner radius, cavity opening angle, and cavity density), and to the accretion disk (mass, accretion rate, outer radius, inner radius, flaring power, and scale height). By considering 10 different viewing angles, the final total grid consists of 200,000 SED.\\
 The RB07 tool requires in input the minimum and maximum distances and, for NGC~3603~YC, we considered 6 and 8 Kpc. To fit the observed data, the RB07 tool requires at least three fluxes and beyond these it accepts up to  N-1 upper/lower limits, where N is the number of the fluxes. In our case, we considered the flux of the multiple associations as upper limit and we obtained reliable comparison for 35 detections. 
Note that the wide range of the theoretical parameters provides degenerate models for each source.
To reduce the degeneracy we imposed a validity range for the theoretical bolometric luminosity ($L_{bol}^T$) calculated by the RB07 tool, by evaluating the observed bolometric luminosity ($L_{bol}^O$) at the minimum (6 Kpc) and maximum (8 Kpc) adopted distances. 
In order to estimate the physical parameters of each source, we provided a statistical analysis of all the selected models in the validity range (Fig.\ref{f7}).
Specifically, we fitted with a Gaussian function the histograms of the parameters and we calculated the mean and its standard deviation. 
The results of this analysis are listed in Tab.\ref{sed}. We obtained that the sources are in the bolometric luminosity range $0.8~10^3 < L_{bol}^T/L_{\sun} \lessapprox 4~10^4$ and that $\sim 70$\% of them has $ L_{bol}^T < 10^{4} L_{\sun}$.  The ranges of the envelope mass ($0.1~10^2 < M_{env}/ M_{\Sun} < 28 ~10^2$) and of the central mass (8 < $ M_{\star}/ M_{\Sun} \lessapprox 27$) suggest that the main part of the original material has already fallen onto the central object, but accretion processes are still ongoing ($ 0.2~10^{-3} < \dot{M}/(M_{\Sun}/yr) < 6.2~10^{-3} $). The central masses also suggest that high mass stars ($M_{\star}/M_{\sun}$ > 8) are forming. 
For $\sim 80$\% of the sources, we found a central object temperature of  $ T \sim 4~10^3$~K  and an age lower than $3.2~10^3$~yr. Only six sources (Her-38, Her-49, Her-59, Her-78, Her-105, Her-106) are hotter ($T > 22~10^3$~K) and older (age $ > 4.9~10^3$~yr). \\
In order to study the spatial distribution of the mass of the central objects, we divided the sample in two different groups for the values above and below $16~M_{\sun}$ (see panel b) of  Fig.\ref{f6}). 
North of the cluster, we found that  objects with higher central mass lie near MM8, MM11 and MM12 clumps. Note that MM11 is also a region of intense maser activity as seen by B10 (see the discussion on Sect.\ref{maser1}).     
South of the cluster, objects with higher central mass well trace the MM1, MM2, MM3 and MM4 clumps. In particular, associated with the MM1 and MM2 pillar heads we found two of the hotter and older massive stars (Her-49, Her-78). 

We investigated the evolutionary phases of the fitted sources by using the $L_{bol}^O-M_{env}$ diagram and the theoretical tracks obtained by \citet{mol08} with the McKee \& Tan (2003) model (see Fig.\ref{f9}).
The evolutionary time of each track starts from the point of maximum $M_{env}$ and at first proceeds almost vertically in the diagram, describing the phase of pre-Zero Age Main Sequence (ZAMS) protostar, where the central object rapidly increases its luminosity.
The track shows a knee when the object approaches to the ZAMS and then it proceeds almost horizontally experiencing the envelope clean-up phase.
We found that 28 sources are located in the pre-ZAMS region and they are candidate Class~0 protostars (see Tab.\ref{sed}). 
The remaining sources are Class~I candidates: 5 sources (Her-10, Her-24, Her-25, Her-49, Her-78) are in the locus of the ZAMS  and 2 sources (Her-105, Her-106) are in the envelope clean-up phase.\\
 Expanding the work by \citet{nurn03b,nurn08} and \citet{Veo10} on the protostellar candidates IRS9A, IRS9B and IRS9C, our results clearly show the presence of a large number of protostellar candidates in areas around NGC~3603~YC, where signatures of star formation activity have been found previously (N02, NS03, B10). 
Owing to the early evolutionary phases of these objects, we concluded that they formed after NGC~3603~YC ($\sim$ 1~Myr). 
These results support the theoretical scenario for which the energy produced by the massive stellar cluster triggered the ongoing star formation (N02, P11).

\subsection{Comparison between Herschel source detections and previous observations}
\label{maser1}

In the investigation provided by B10, the authors used the Australia Telescope Compact Array (ATCA) to measure with arcsecond accuracy the positions for 379 masers at the 22-GHz transition of water.  The principal observation targets were 202 OH masers associated with star forming regions in the Southern Galactic plane, while in a second epoch new targets of methanol masers were added. Among them, 6 masers are located in the regions of NGC~3603~YC \citep[see also][]{caswell89,caswell04}, as plotted in panel a) of Fig.\ref{f6}. In order to investigate the far-IR Herschel counterparts, we took into account the uncertainty on the Herschel source positions, given by the FWHM of the shortest wavelength in which the source was measured (see Tab.1).

The result of the match shows that 4 masers are well detected within the position uncertainty, as listed in Tab.\ref{maser} (and also marked in Tab.2 with the letter $B$). The three masers associated with the two prominent pillars close to NGC~3603~YC are also well detected. In particular, B-4 and B-5 are associated with the Her-49 and Her-78 sources, respectively,  which are between the older ($\sim 26-39 ~10^3$~yr) and more evolved (Class~I) objects, as described in Sect.\ref{pointsed}.  

By using the TIMMI2@3.6 ESO telescope, NS03 obtained a mid-IR (at 11.9  and 18 $\mu$m) survey of the NGC~3603 region and detected 36 point sources and 42 diffuse emission knots. To find their Herschel counterparts, we used the maximum distance of 10\farcs that is related to the resolution and pixel size of the two instruments.   We found that 11 objects strictly satisfy our condition, among these 7 are considered by NS03 as point sources and 4 diffuse emission knots. The common detections are listed in Tab.\ref{maser} and marked with  {\it NS} in Tab.2. 
We found Herschel counterparts for the NS-1 and NS-2 sources, likely associated to the maser emission near the MM11 star forming region (NS03).  While it was not possible to fit the NS-2 counterpart (also related to B-2 maser),  the NS-1 source is associated to Her-59, which represents a YSO ($\sim 21.5 M_{\Sun}$ , $\sim 5~10^3$ yr).  
In MM11 we also detected the NS-15 diffuse emission that NS03 found related to a methanol maser. We found that the last source corresponds to the Her-68 detection, fitted by a massive YSO ($\sim 18.5 M_{\Sun}$, $\sim 10^3$ yr). 

The source NS-4 is associated to IRS4, which is a supergiant of spectral type MIb \citep{frogel}, very bright in the mid-IR bands. For its Herschel counterpart, we can not discriminate between two detections (Her-55, Her-56) that belong to a multiple association and we decided to list both options.
The source NS-9A is located near the MM2 clump, and it is described by NS03 as young protostar with age $<10^5$ yr. Although we found its Herschel counterpart, the crowding (multiple association) prevents the possibility to reconstruct the SED.  However,  being the brightest source of this region at 11.9 and 18  $\mu$m, NS-9A also appears as the brightest component of the multiple association in the 21.3 MSX and 70 $\mu$m Herschel bands. 
The NS-12 source is supposed to be a young star toward the very center of NGC~3603~YC \citep[see also][]{tapia01}; owing to its position near the multiple association formed by Her-64 and Her-65, its counterpart is not clear and we listed both detections. 
The NS-13A and NS-13B were associated to the water maser detected in the MM4 clump. 
Both sources are associated to same Herschel counterpart (Her-87), which we fitted with a YSO model of $\sim 2.9~10^3$~yr and 19.6~$M_{\Sun}$.
We also found Herschel counterparts for the knots NS-28 and NS-36 located around the MM2 clump and for the NS-48 located near MM4, but we have no SED results for these objects.

\section{Summary and Conclusions}
\label{conclusion}
We investigated the physical properties of the cloud complex surrounding NGC~3603~YC in the far-IR (70-500 $\mu$m) PACS and SPIRE Herschel bands. In order to study the diffuse cold dust component we provided the temperature and the column density maps at high resolution by using the 70-160 $\mu$m flux ratio. Moreover, we checked our results by using the Herschel bands not affected by VSG IR emission. We found that the average temperature of the central regions (36~K), the column density ($log_{10}N_H$=22.0~cm$^{-2}$) and the visual extinction ($A_V=5.3$ mag) are in good agreement with the optical and near-IR literature estimates. 
 We also provided a photometric analysis of the Herschel images obtaining 107 point-like sources.
By including 4 MSX and 2 WISE images, we were able to compare the observed SED with the YSO models for  35 detected sources. 
We found that the main part ($\sim 80$\%) of the fitted sources have an age $<~3.2~10^3$ yr.  Moreover, we found that 28 objects belong to Class~0 and the remaining to Class~I.
  Beyond the initial studies of the protostellar candidates IRS9A, IRS9B and IRS9C by \citet{ nurn03b, nurn08} and \citet{Veo10}, our results significantly increase the number of very young sources in the HII region around the NGC~3603~YC.  The early evolutionary phases of these objects confirm an ongoing star formation activity, as previously suggested by the maser and mid-IR observations (N02, NS03, P11). 
Taking into account the age of the NGC~3603~YC ($\sim$1~ Myr), this investigation supports the scenario of  a sequential star formation in its surrounding, where the birth of younger stars is likely triggered by the strong radiation field emitted by the massive stars in NGC~3603~YC.

\begin{table*}
\begin{center}
\caption{Herschel, MSX and WISE instrumental performance} 
\begin{tabular}{|c|c|c|c|c|c|c|c|c|c|}

\hline
 Herschel [$\mu$m] & FWHM [\farcs] & sensitivity limit [mJy]  & pixel scale [\farcs /pxl] \\
 \hline 
 70  & 5       & 3.0   &  3.2\\
 160   & 13    & 5.0    &  4.5 \\
250    &  18     & 6.4    &  6.0 \\
 350 &  25   & 5.3  & 8.0    \\
 500    &  36    & 7.7    & 11.5  \\

\hline
MSX [$\mu$m] & & & \\  
\hline
 8.3   & 20    & 1.9   & 6 \\
 12.1   & 20      &  4.7   &  6 \\
14.7   & 20    & 3.6  &   6  \\
 21.3    &  20     & 11.9   & 6 \\
 \hline
WISE [$\mu$m] & & & \\ 
\hline
3.4 & 6  & 0.07 & 1.38 \\
4.6 & 6.8 & 0.1 & 1.38 \\
\hline 
 \end{tabular}
 \label{tab1}
\end{center}
\end{table*}

\newpage
\footnotesize{
\begin{landscape}
\relsize{-1}
\begin{longtable}{|c|c|c|c|c|c|c|c|c|c|c|c|c|c|}
\caption{Sources detected by Herschel in the N02 region and their MSX counterparts. The WISE measurements are only for the detections in common between Herschel and MSX. The $U$ letter marks an upper limit flux (calculated via Eq.4), while the $\dagger$ symbol accounts the sources fitted with a SED model. The counterparts of the B10 and NS03 sources are marked with $B$ and $NS$ respectively.}  \\
\hline
\multicolumn{3}{|c|}{ } & \multicolumn{2}{c|}{ \small{ WISE }} & \multicolumn{4}{c|}{ \small{MSX }} & \multicolumn{5}{c|}{ \small {Herschel}} \\ [6pt] 
\hline
Her-ID & Ra & Dec & $3.4\mu$m &$4.6\mu$m & $8.3\mu$m &$12.1\mu$m &$14.7\mu$m & $21.3\mu$m & $70\mu$m & $160\mu$m & $250\mu$m & $350\mu$m & $500\mu$m \\
\hline
 & hh:mm:ss & dd:mm:ss &  Jy & Jy &  Jy & Jy & Jy & Jy & Jy & Jy & Jy & Jy &  Jy \\
\endhead
 \hline
1$^\dagger$ &  11:14:38.85 & -61:15:33.2 &    -     &    -     &    0.04$^U$     &    0.05$^U$     &    0.06$^U$     &    0.25 &   14.38 &   34.26 &   15.68 &    -     &    -     \\
\hline
\hline
2  &  11:14:40.86 &  -61:16:00.3  &  -     &    -     &   0.07 &    -     &    -     &    -     &   47.50 &  36.60 & 10.92 &    -     &    -    \\
 \hline
\hline
3$^\dagger$ & 11:14:41.05 & -61:19:10.5 &    -     &    -     &    0.03 &    0.03$^U$     &    0.04$^U$     &    0.15$^U$     &   36.35 &   28.89 &   14.51 &    7.56 &    -      \\
\hline
4 &   11:14:42.99 &  -61:08:25.3 &    -     &    -     &    -     &    -     &    -     &    -     &    -     &   23.94 &   44.84 &   25.35 &    9.15 \\
\hline
\hline
5 &  11:14:43.87 & -61:13:19.7 &    -     &    -     &    -     &   -     &    0.26 &    -     & 19.99 & 33.02 & 27.46 &  40.76 & \multirow{2}{*}{13.86} \\
\cline{1-13}
6$^\dagger$ &  11:14:47.66 & -61:12:44.8 &    -     &    -     &    0.08 &    -     &    0.06$^U$     &    0.27$^U$     &   42.105 &   49.00 &    -     &   10.78 &  \\
\hline
\hline
7 &   11:14:44.47 & -61:16:51.2 &    -     &    -     &    -     &    -     &    -     &    -     &   24.14 &   34.18 &    -     &    -     &    -    \\
\hline
8 &  11:14:46.26 & -61:10:37.9 &    -     &    -     &    -     &    -     &    -     &    -     &    -     &    -     &   22.84 &    8.99 &    -    \\
\hline
9 &   11:14:46.44 &  -61:06:20.6 &    -     &    -     &    -     &    -     &    -     &    -     &    -     &    -     &   14.12 &   16.76 &    -    \\
\hline
10$^\dagger$ &  11:14:47.62 & -61:16:37.2 &    0.02     &    0.05     &    0.17 &    0.13 & -         &    0.63 &  131.93 &   47.22 &    -     &    -     &    -  \\
\hline
11$^\dagger$ &  11:14:46.71 & -61:12:12.5 &    -     &    -     &    0.03 &    0.05$^U$     &    0.06$^U$     &    0.20$^U$     &   41.34 &   30.45 &    8.43 &    2.48 &    -    \\
\hline
\hline
12 &   11:14:48.64 &  -61:07:19.2 &    -     &    -     &    -     &    -     &    -     &    -     &    -     &   24.09 &   16.0 &   \multirow{2}{*}{18.87} &   \multirow{2}{*}{14.29} \\
\cline{1-12}
13 &   11:14:49.87 &  -61:07:39.6 &    -     &    -     &    -     &    -     &    -     &    -     &    -     &   19.72 &   22.01 &    &    \\
\hline
\hline
14$^\dagger$   &  11:14:48.76 &  -61:17:02.7 &    0.03     &    0.14     &    0.17 &    0.19$^U$     &    0.29$^U$     &    0.88$^U$     &  215.53 &  132.34 &   30.54 &   16.49 &    4.30 \\
\hline
15 &   11:14:49.02 & -61:17:42.0 &    -     &    -     &    -     &    -     &    -     &    -     &   22.61 &   16.46 &    -     &    -     &    -  \\
\hline
\hline
16 &   11:14:51.11 & -61:15:12.2 &    -     &    -     &    -     &    -     &    -     &    -     &   34.97 &   \multirow{2}{*}{32.03} &   \multirow{2}{*}{28.92} &    -     &    -     \\
\cline{1-10}\cline{13-14}
17 &   11:14:52.66 &  -61:15:00.3 &    -     &    -     &    -     &    -     &    -     &    4.51 &   55.68 &    &    &    -     &    -     \\
\hline
\hline
18$^\dagger$   &  11:14:51.17 & -61:10:14.3 &    -     &    -     &    -     &    -     &    0.02$^U$     &    0.09$^U$     &    53.09$^U$     &    82.11     &   22.18 &   25.73 &   24.89  \\
\hline
\hline
19 &   11:14:51.62 & -61:13:32.5 &    -     &    -     &    -     &    -     &    -     &    -     &    -     &   64.20 &    9.15 &   \multirow{3}{*}{80.01} &   \multirow{3}{*}{53.68} \\
\cline{1-12}
20 &    11:14:53.79 & -61:13:51.3 &    -     &    -     &    -     &    -     &    -     &    -     &  224.18 &  135.03 &  \multirow{2}{*}{115.06} &   &    \\
\cline{1-11}
21 &   11:14:53.81 & -61:13:29.0 &    -     &    -     &    -     &    -     &    -     &    -     &    -     &   69.27 &   &    &   \\
\hline
\hline
22 & 11:14:52.71 & -61:15:42.4 &    -     &    -     &    -     &    -     &    -     &    -     &  143.32 &  \multirow{2}{*}{66.02} &   \multirow{2}{*}{36.89} &   \multirow{2}{*}{11.98} &    - \\
\cline{1-10}\cline{14-14}
23 & 11:14:53.58 & -61:15:37.5 &    -     &    -     &    -     &    -     &    -     &    -     &  148.55 &    &    &   &    -  \\
\hline
\hline
24$^\dagger$ &  11:14:53.73 & -61:13:18.5 &    -     &    -     &    0.18 &    0.25 &    0.39 &    1.99 &  118.54 &    73.93$^U$     &    22.18$^U$     &    -     &    -     \\
\hline
25$^\dagger$  &  11:14:54.19 &  -61:20:3.8 &    -     &    -     &    -     &    -     &    0.02$^U$     &    0.08$^U$     &   22.67 &   15.42 &    4.47 &    -     &    -      \\
\hline
\hline
26 &  11:14:54.36 & -61:14:30.7 &    -     &    -     &    -     &    -     &    -     &    -     &   31.32 &   \multirow{2}{*}{11.37} &    -     &    -     &    -     \\ 
\cline{1-10}\cline{12-14}
27 &  11:14:55.62 & -61:14:31.3 &    -     &    -     &    -     &    -     &    -     &    -     &   41.45 &    &    -     &    -     &    -     \\
  
\hline
\hline
28 & 11:14:54.73 & -61:12:52.3 &    -     &    -     &    -     &    -     &    -     &    -     &   73.29 &   \multirow{2}{*}{96.15} &   \multirow{2}{*}{45.50} &   \multirow{2}{*}{44.09} &   \multirow{2}{*}{41.72} \\
\cline{1-10}
29 &  11:14:57.82 & -61:12:54.2 &    -     &    -     &    -     &    -     &    -     &    -     &  153.88 &    &    &    &    \\
\hline
\hline
30 &  11:14:55.28 & -61:13:11.6 &    -     &    -     &    -     &    0.20 &    -     &    -     &   92.54 &    -     &    -        &    -     &  -  \\
\hline
31 &   11:14:56.09 & -61:12:54.5 &    -     &    -     &    0.17 &    -     &    -     &    -     &  224.11 &    -     &    -         &    -     &  -  \\
\hline
32 &  11:14:56.31 & -61:13:58.6 &    -     &    -     &    0.30 &    -     &    -     &    -     &   41.45 &    -     &    -       &    -     &  -  \\
\hline
\hline
33$^\dagger$ &  11:14:57.08 &  -61:09:55.3 &    -     &    -     &    -     &    -     &    0.02$^U$     &    0.06$^U$     &    87.50     &   40.88 &   15.97 &   10.80 &   \multirow{2}{*}{49.57}  \\
\cline{1-13}
34$^\dagger$  &  11:14:59.01 & -61:10:33.5 &    -     &    -     &    -     &    -     &    0.04$^U$     &    0.15$^U$     &    156.27     &   84.68 &   63.91 &   49.49 &    \\
\hline
\hline
35 &  11:14:57.70 & -61:13:17.6 &    -     &    -     &    -     &    -     &    -     &    -     &   46.95 &   13.95 &    -     &    -     &    -     \\
\hline
36$^\dagger$ &  11:14:58.36 &   -61:08:04.1 &    0.02     &    0.04     &    0.07 &   -   &    -   &    0.02$^U$     &   126.67 &  103.33 &   54.33 &   33.18 &   31.04  \\
\hline
37  &  11:14:58.48 & -61:14:17.0 &    -     &    -     &    -     &    -     &    -     &    -     &   64.09 &   15.94   &    -     &    -     &  -  \\
\hline
38$^\dagger$  &  11:14:59.20 & -61:11:33.9 &    0.08     &    0.19     &    0.36 &    0.17 &    0.19 &    0.99 &  355.08 &  341.81 &  276.08 &  126.54 &   57.36  \\  
\hline
39$^\dagger$   &  11:14:59.23 &  -61:13:03.92 &    -     &    -     &    0.16$^U$     &    0.22 &    0.47 &    0.84$^U$     &   72.71 &   95.01 &    -     &    -     &    -     \\
\hline
\hline
40 &   11:15:00.08 & -61:16:31.0 &    -     &    -     &    -     &    -     &    -     &    -     &  180.22 &  \multirow{2}{*}{192.61} &    -     &    -     &    -     \\   
\cline{1-10}\cline{12-14}
41 &   11:15:00.34 & -61:16:40.8 &    -     &    -     &    -     &    -     &    -     &    -     &  148.64 &   &    -     &    -     &    -     \\ 
\hline
\hline
42$^\dagger$ &   11:15:00.10 & -61:11:55.6 &    0.05     &    0.10     &    0.26 &    -    &    -  &    0.27$^U$     &   90.69 &  124.73 &   50.67 &   22.29 &    -   \\
\hline
43$^\dagger$ &   11:15:00.33 &  -61:08:21.8 &    0.03     &    0.07     &    0.08 &    -     &    -     &    0.20 &   43.78 &   45.28 &   35.98 &    -     &    -       \\
\hline
44 &  11:15:00.42 & -61:10:41.7 &  -     &    -     &   0.13 &      -     &    -     &    -     &   31.36 &    -     &    -     &    -     &    -    \\
\hline
\hline
45 &  11:15:00.54 & -61:13:51.4 &    -     &    -     &    -     &    -     &    -     &    -     &  118.80 &   \multirow{2}{*}{59.06} &   \multirow{2}{*}{19.28} &    -     &    -     \\
\cline{1-10}\cline{13-14}
46 &   11:15:00.86 & -61:13:54.7 &    -     &    -     &    -     &    -     &    -     &    -     &   82.23 &    &    &    -     &    -     \\
\hline
\hline
47 &   11:15:01.66 &  -61:06:48.3 &    -     &    -     &    -     &    -     &    -     &    -     &    -     &    -     &   11.51 &    5.65 &     -  \\
\hline
\hline
48$^\dagger$ &   11:15:01.67 & -61:16:25.9 &    0.04     &    0.08     &    0.58 &    1.15 &    1.44 &    4.80$^U$     &  509.85 &  348.53 &  114.00 &   53.70 &   \multirow{2}{*}{24.12} \\          
\cline{1-13}
49$^{\dagger,B}$  &   11:15:02.47 & -61:15:53.1 &    0.15     &    1.06      &    4.41 &    1.76 &    7.37 &   27.24 &  561.02 &  372.38 &  124.87 &   43.01 &  -    \\
\hline
\hline
50 &    11:15:02.25 & -61:13:20.2 &    -     &    -     &    -     &    -     &    -     &    -     &   98.18 &  \multirow{2}{*}{144.89} &  \multirow{2}{*}{105.83} &   \multirow{2}{*}{52.48} &   \multirow{2}{*}{54.73} \\
\cline{1-10}
51 &   11:15:03.06 & -61:13:15.0 &    -     &    -     &    -     &    -     &    -     &    -     &   63.37 &   &   &    &    \\
\hline
\hline
52$^\dagger$  &   11:15:02.48 &  -61:19:05.6 &    -     &    -     &    0.14 &    -     &    -     &    0.79$^U$     &    515.52$^U$     &    306.24$^U$     &   46.14 &   33.88 &   31.85 \\
\hline
\hline
53 &   11:15:02.68 &  -61:12:10.0 &    -     &    -     &    -     &    -     &    -     &    -     &  527.66 &   \multirow{2}{*}{84.34} &    -     &    -     &    -  \\          
\cline{1-10}\cline{12-14}
54 &   11:15:02.80 & -61:12:11.2 &    -     &    -     &    -     &    -     &    -     &    -     &   93.99 &    &    -     &    -     &    -    \\
\hline
\hline
55$^{NS}$ &   11:15:03.23 & -61:14:21.6 &    -     &    -     &    0.24 &    -     &    -     &    -     &   60.23 &   \multirow{2}{*}{24.22} &   \multirow{2}{*}{11.79} &    -       &  -  \\
\cline{1-10}\cline{13-14}
56$^{NS}$ &   11:15:04.17 & -61:14:25.0 &    -     &    -     &    -     &    -     &    -     &    -     &   47.24 &    &    &    -     &    -     \\
\hline
\hline
57 &   11:15:03.96 &  -61:08:47.8 &    -     &    -     &    -     &    -     &    -     &    -     &    -     &   17.50 &   17.72 &    -     &    -   \\
\hline
58$^\dagger$ &   11:15:04.19 &  -61:18:05.3 &    -     &    -     &    0.41 &    -     &    0.19$^U$     &    0.79$^U$     &   71.03 &   32.33 &    -     &    -     &    -     \\
\hline
\hline
59$^{\dagger,NS}$ &   11:15:04.28 &  -61:09:36.1 &    -     &    0.05     &    0.08 &    0.030 &    0.03 &    0.33 &  653.30 & \multirow{2}{*}{1382.02} &  \multirow{2}{*}{659.69} &  \multirow{2}{*}{425.70} &  \multirow{3}{*}{188.23}  \\
\cline{1-10}
60$^{B, NS}$  &   11:15:06.00 &  -61:09:38.8 &    -     &    -     &    -     &    -     &    -     &    -     &  667.03 &  &   &   &   \\
\cline{1-13}
61   &   11:15:05.21 &  -61:09:28.8 &    -     &    -     &    -     &    -     &    -     &    -     &    -     &    -     &    -     &  141.58 &   \\
\hline
\hline
62  &   11:15:04.39 &  -61:19:00.6 &    -     &    -     &    -     &    0.04 &    -     &    -     &   23.50 &   38.54 &    -     &    -     &    -    \\
\hline
\hline
63  &   11:15:04.86 & -61:17:55.5 &    -     &    -     &    -     &    -     &    -     &    -     &   65.02 &  \multirow{2}{*}{105.58} &  \multirow{3}{*}{158.64} &   \multirow{3}{*}{69.87} &    -   \\
\cline{1-10}\cline{14-14}
64$^{NS}$  &   11:15:07.12 & -61:18:00.5 &    -     &    -     &    -     &    -     &    -     &    -     &   18.65 &   &   &    &    -    \\
\cline{1-11}\cline{14-14}
65$^{NS}$  &   11:15:07.08 &  -61:17:52.0 &    -     &    -     &    -     &    -     &    -     &    -     &   84.01 &  127.54 &   &    &    -   \\
\hline
\hline
66 &  11:15:05.05 & -61:13:14.7 &    -     &    -     &    -     &    -     &    -     &    -     &    -     &  104.68 &   \multirow{2}{*}{58.02} &    -     &    -      \\       
\cline{1-11}\cline{13-14}
67 &    11:15:06.50 &  -61:13:04.0 &    -     &    -     &    -     &    -     &    -     &    -     &  132.04 &  110.37 &    &    -     &    -      \\
\hline
\hline
68$^{\dagger,NS}$ &    11:15:06.4 &  -61:10:04.92 &    0.02     &    0.03     &    0.03 &    0.01$^U$     &    0.01$^U$     &    0.13 &  142.75 &  567.45 &  255.29 &    -     &    -       \\
\hline
69 &   11:15:07.72 & -61:16:31.7 &    -     &    -     &    -     &    2.34 &    -     &    -     &   34.67 &    -     &    -     &    -     &    -     \\
\hline
70 &   11:15:07.77 &  -61:11:07.8 &    -     &    -     &    -     &    -     &    -     &    -     &    -     &   38.86 &   25.99 &    -     &    -     \\
\hline
\hline
71 &  11:15:07.90 &  -61:17:07.2 &    -     &    -     &    -     &    -     &    -     &    -     &   58.44 &    -     &  \multirow{2}{*}{255.47} &   \multirow{2}{*}{56.86} &    -   \\
\cline{1-11}\cline{14-14}
72$^{B}$ &   11:15:08.51 & -61:16:57.0 &    -     &    -     &    -     &    -     &    -     &    -     &  425.58 &    -     &   &    &    -   \\
\hline
\hline
73 &    11:15:08.26 &  -61:14:07.2 &    -     &    -     &    -     &    -     &    -     &    -     &  150.65 &   \multirow{3}{*}{82.37} &   \multirow{4}{*}{34.61} &   \multirow{4}{*}{12.43} &    \multirow{4}{*}{8.08}  \\
\cline{1-10}
74 &   11:15:08.69 &  -61:14:08.3 &    -     &    -     &    -     &    -     &    -     &    -     &  103.42 &    &    &    &     \\
\cline{1-10}
75 &   11:15:10.32 & -61:14:06.1 &    -     &    -     &    -     &    -     &    -     &    -     &   28.37 &    &    &    &    \\
\cline{1-11}
76 &   11:15:10.16 &  -61:14:24.7 &    -     &    -     &    0.21 &    -     &    -     &    -     &   39.24 &   23.62 &    &    &   \\
\hline
\hline
77$^{NS}$ &   11:15:09.11 & -61:20:54.5 &    -     &    -     &    0.14 &    -     &    -     &    -     &   46.07 &    -     &    -     &    -     &    -     \\
\hline
78$^{\dagger,B}$   &   11:15:09.13 & -61:16:24.1 &    0.14     &    0.22     &    2.10 &    1.42 &    2.79 &   13.44 &    608.21     &  229.99 &  117.03 &   50.74 &    -    \\
\hline
79 &    11:15:09.32 & -61:11:47.2 &    -     &    -     &   0.05 &    -     &    -     &    -     &   25.92 &     -     &    -   & - & -   \\
\hline
\newpage
\hline
80 &  11:15:09.45  & -61:16:58.0 &    -     &    -     &   2.43  &   4.24 & 3.81 & 8.32 &    -     & 536.38 & \multirow{2}{*}{388.70} & \multirow{2}{*}{279.21} & \multirow{4}{*}{197.87}   \\ 
\cline{1-11}
81$^{NS}$ &  11:15:09.96 & -61:16:45.9 &    -     &    -     &    -     &    -     &    -     &   15.97 &  475.50 &  423.91 &   &   &   \\
\cline{1-13}
82 &   11:15:11.28 & -61:17:15.3 &    -     &    -     &    -     &    -     &    -     &    -     &  137.72 &  369.00 &  225.52 &  \multirow{2}{*}{145.62} &   \\            
\cline{1-12}
83$^\dagger$   &  11:15:14.48 & -61:17:26.6 &    -     &    0.05     &    -     &    -     &    -     &    0.91$^U$     &  161.89 &  276.97 &  124.28 &   &  \\
\hline
\hline
84 &    11:15:09.50 &  -61:12:08.1 &    -     &    -     &    -     &    -     &    -     &    -     &   28.08 &   16.43 &    -     &    -     &    -     \\
\hline
85$^\dagger$  &   11:15:09.63 & -61:11:36.6 &    -     &    -     &    -     &    -     &    0.18$^U$     &    0.75$^U$     &    314.62     &   96.43 &   40.35 &   13.93 &   13.03    \\
\hline
86$^\dagger$ &  11:15:10.42 & -61:21:51.4 &    -     &    -     &    -     &    0.02$^U$     &    0.01$^U$     &    0.06$^U$     &    46.07     &   38.00 &   20.38 &   11.56 &    -       \\
\hline
87$^{\dagger,NS}$  &  11:15:10.85 & -61:20:31.1 &    -     &    0.02     &    0.04$^U$    &    0.02$^U$     &    0.02$^U$     &    0.07$^U$     &  447.18 &  492.61 &  316.13 &  121.82 &   72.03   \\
\hline
88 &   11:15:11.45 & -61:19:29.7 &    -     &    -     &    -     &    -     &    -     &    -     &   36.94 &  114.04 &   67.71 &    -     &    -   \\
\hline
89 &  11:15:11.59 & -61:10:21.6 &    -     &    -     &    -     &    -     &    -     &    -     &    -     &   19.57 &   53.98 &    -     &    -   \\
\hline
\hline
90 &  11:15:11.78 & -61:18:21.8 &    -     &    -     &    -     &    -     &    -     &    -     &    -     &    -     &    -     &   96.10 &   \multirow{3}{*}{46.52} \\            
\cline{1-13}
91$^\dagger$  &  11:15:12.43 & -61:18:42.2 &    -     &    0.03     &    0.09 &    -     &    0.13$^U$     &    0.46$^U$     &   44.33 &  119.20 &   67.46 &    -  &      \\
\cline{1-13}
92$^\dagger$ &  11:15:12.66 & -61:18:50.8 &    0.01     &    0.02     &    -     &    -     &    0.03$^U$     &    0.15$^U$     &   71.52 &  180.29 &   65.41 &   53.46 &    \\
\hline
\hline
93$^{NS}$ &  11:15:14.07 & -61:16:57.1 &    -     &    -     &    -     &    2.14 &    -     &    -     &    -     &    -     &   55.39 &    -     &    -    \\
\hline
\hline
94 &  11:15:15.01 & -61:16:15.9 &    -     &    -     &    -     &    -     &    -     &    -     &  115.68 &   \multirow{2}{*}{28.49} &    -     &    -     &    -      \\            
\cline{1-10}\cline{12-14}
95$^{NS}$ &  11:15:17.09 &  -61:16:05.1 &    -     &    -     &    -     &    -     &    -     &    -     &   47.51 &    &    -     &    -     &    -     \\            
\hline
\hline
96 &  11:15:15.51 & -61:20:18.8 &    -     &    -     &    -     &    -     &    -     &    -     &    -     &   66.64 &   98.48 &    -     &    -     \\ 
\hline
\hline
97   &  11:15:16.32 & -61:10:50.4 &    -     &    -     &    -     &    -     &    -     &    -     &    -     &   72.11 &   68.88 &   \multirow{3}{*}{58.50} &   \multirow{3}{*}{40.44}  \\
\cline{1-12}
98$^\dagger$  &  11:15:18.85 & -61:10:56.8 &    0.02     &    0.09     &    0.16 &    -     &    -     &    0.45$^U$     &   19.22 &  146.30 &   37.11 &    &   \\
\cline{1-12}
99$^\dagger$  &  11:15:22.02 &  -61:11:02.9 &    -     &    -     &    0.19 &    -     &    -     &    0.55$^U$     &   32.43 &  114.90 &   95.52 &    &      \\
\hline
\hline
100 &   11:15:18.85 & -61:16:56.0 &    -     &    -     &    -     &    -     &    -     &    -     &   36.69 &   \multirow{2}{*}{26.73} &    -     &    -     &    -      \\            
\cline{1-10}\cline{12-14}
101 &  11:15:19.58 & -61:16:43.0 &    -     &    -     &    -     &    -     &    -     &    -     &   22.72 &    &    -     &    -     &    -      \\
\hline
\hline
102$^\dagger$ &  11:15:19.46 & -61:20:17.7 &    -     &    -     &    0.07 &    -     &    0.03$^U$     &    0.12$^U$     &   37.47 &   36.02 &    -     &    -     &    -        \\
\hline
103 &  11:15:20.72 & -61:18:10.8 &    -     &    -     &    -     &    -     &    -     &    -     &   22.96 &    4.85 &    -     &    -     &    -   \\
\hline
104$^\dagger$&   11:15:22.35 & -61:17:46.0 &    0.01     &    0.04     &    0.05 &    -     &    0.09$^U$     &    0.32$^U$     &   33.45 &   19.19 &    -     &    -     &    -      \\
\hline
105$^\dagger$  &  11:15:22.58 & -61:17:13.0 &    -     &    -     &    0.35 &    0.63$^U$     &    0.36 &    1.24 &   37.55 &   44.19 &    30.00$^U$     &    -     &    -     \\
\hline
106$^\dagger$ &  11:15:24.45 & -61:15:59.1 &    0.01     &    0.03     &    0.13 &    0.37 &    0.30 &    3.25 &   48.99 &    34.96$^U$     &    11.15$^U$     &    -     &    -       \\
\hline
107 &  11:15:28.58 & -61:10:11.9 &    -     &    -     &    -     &    -     &    -     &    -     &    -     &    -     &   38.81 &   65.02 &   54.45 \\
\hline
\end{longtable}
\label{tab2}
\end{landscape}  
 }

\newpage
\begin{deluxetable}{ccccccccccc} 
\tabletypesize{\scriptsize}
\tablecaption{SED analysis physical parameters}
\label{sed} 
\tablehead{
\colhead{Her-ID}&
\colhead{L$_{bol}^0$ \tablenotemark{a}}&
\colhead{L$_{bol}^T$}&
\colhead{M$_{env}$}&
\colhead{$\dot{M}$}&
\colhead{M$_\star$}&
\colhead{T$_\star$}&
\colhead{Age}&
\colhead{Class}\\
\colhead{}&
\colhead{ $10^3~L_\sun$ }&
\colhead{$10^3~L_\sun$ }&
\colhead{$10^3M_{\sun}$}&
\colhead{$10^{-3}~M_{\sun}~yr^{-1}$}&
\colhead{$M_{\sun}$ }&
\colhead{$10^3~K$ }&
\colhead{$10^3~yr$ }&
\colhead{}}
\startdata
  1 &	1.20$\pm$0.50 &   0.88$\pm$  0.02&  0.08$\pm$ 0.03 &  0.41$\pm$ 0.09 &  8.2$\pm$ 0.2   &  4.25$\pm$ 0.08  &   2.20$\pm$  0.96  & 0 \\ 
  3 &	2.06$\pm$0.55 &   2.15$\pm$  0.26&  0.27$\pm$ 0.14 &  1.16$\pm$ 0.28 &  9.6$\pm$ 0.5   &  4.32$\pm$ 0.08  &   2.31$\pm$  0.95  & 0 \\ 
  6 &  2.86$\pm$0.80 &   3.08$\pm$  0.46&  0.40$\pm$ 0.16 &  0.95$\pm$ 0.22 & 10.4$\pm$ 0.6   &  4.23$\pm$ 0.06  &   1.26$\pm$  0.25  & 0 \\ 
 10 &  5.56$\pm$1.55 &   5.46$\pm$  0.85&  0.14$\pm$ 0.04 &  0.32$\pm$ 0.07 & 10.7$\pm$ 0.9   &  3.96$\pm$ 0.15  &   2.94$\pm$  0.56  & I \\ 
 11 &	2.23$\pm$0.65 &   2.14$\pm$  0.28&  0.19$\pm$ 0.10 &  0.98$\pm$ 0.21 &  9.6$\pm$ 0.6   &  4.32$\pm$ 0.08  &   1.72$\pm$  0.35  & 0 \\ 
 14 &  10.91$\pm$2.95 &  10.70$\pm$  1.60&  0.80$\pm$ 0.16 &  1.72$\pm$ 0.28 & 16.0$\pm$ 1.7   &  4.10$\pm$ 0.07  &   0.95$\pm$  0.24  & 0 \\ 
 18 &	3.71$\pm$1.10 &   3.84$\pm$  0.49&  1.40$\pm$ 0.37 &  3.30$\pm$ 0.69 & 11.9$\pm$ 0.1   &  4.25$\pm$ 0.06  &   1.29$\pm$  0.31  & 0 \\ 
 24 &	6.88$\pm$2.90 &   5.12$\pm$  0.51&  0.09$\pm$ 0.05 &  0.28$\pm$ 0.10 &  9.3$\pm$ 0.3   &  4.00$\pm$ 0.17  &   3.17$\pm$  0.98  & I \\ 
 25 &	 1.14$\pm$0.40 &   0.98$\pm$  0.13&  0.05$\pm$ 0.02 &  1.04$\pm$ 0.18 &  8.0$\pm$ 0.7   &  4.27$\pm$ 0.07  &   3.09$\pm$  0.99  & I \\ 
 33 &	3.76$\pm$1.15 &   3.67$\pm$  0.52&  0.60$\pm$ 0.23 &  1.99$\pm$ 0.32 & 11.3$\pm$ 0.7   &  4.18$\pm$ 0.03  &   1.31$\pm$  0.26  & 0 \\ 
 34 &	7.49$\pm$2.25 &   7.71$\pm$  0.96&  1.62$\pm$ 0.38 &  3.45$\pm$ 0.62 & 14.8$\pm$ 0.2   &  4.18$\pm$ 0.04  &   1.46$\pm$  0.23  & 0 \\ 
 36 &  6.19$\pm$2.25 &   4.05$\pm$  0.10&  0.97$\pm$ 0.23 &  1.85$\pm$ 0.31 & 18.2$\pm$ 0.7   &  4.20$\pm$ 0.09  &   0.99$\pm$  0.29  & 0 \\ 
 38 &  22.75$\pm$6.85 &  25.35$\pm$  3.16&  0.87$\pm$ 0.14 &  1.62$\pm$ 0.28 & 27.2$\pm$ 0.9   & 29.76$\pm$ 1.31  &   8.78$\pm$  2.12  & 0 \\ 
 39 &	4.57$\pm$1.30 &   3.94$\pm$  0.53&  0.34$\pm$ 0.13 &  0.77$\pm$ 0.14 & 11.2$\pm$ 0.8   &  4.16$\pm$ 0.09  &   1.24$\pm$  0.75  & 0 \\ 
 42 &	6.25$\pm$1.80 &   5.98$\pm$  0.88&  0.75$\pm$ 0.21 &  1.54$\pm$ 0.32 & 16.1$\pm$ 0.9   &  4.23$\pm$ 0.09  &   3.07$\pm$  1.50  & 0 \\ 
 43 &	2.81$\pm$0.85 &   2.68$\pm$  0.40&  0.16$\pm$ 0.05 &  0.26$\pm$ 0.06 &  9.3$\pm$ 0.6   &  4.18$\pm$ 0.08  &   0.95$\pm$  1.01  & 0 \\ 
 48 &  29.13$\pm$8.65 &  26.85$\pm$  4.00&  0.86$\pm$ 0.13 &  1.58$\pm$ 0.27 & 16.3$\pm$ 1.3   &  3.65$\pm$ 0.22  &   3.13$\pm$  1.26  & 0 \\ 
 49 &  41.72$\pm$12.60 &  40.40$\pm$  6.31&  0.47$\pm$ 0.08 &  0.97$\pm$ 0.21& 20.1$\pm$ 1.2   & 34.11$\pm$ 0.78  &  25.97$\pm$  1.64  & I \\ 
 52 &  23.68$\pm$6.75 &  23.43$\pm$  3.50&  1.57$\pm$ 0.27 &  3.13$\pm$ 0.47 & 16.3$\pm$ 0.9   &  4.07$\pm$ 0.26  &   2.87$\pm$  1.22  & 0 \\ 
 58 &	3.06$\pm$1.40 &   2.09$\pm$  0.04&  0.16$\pm$ 0.09 &  0.60$\pm$ 0.14 & 10.3$\pm$ 0.7   &  4.18$\pm$ 0.08 &   1.22$\pm$  1.10  & 0 \\ 
 59 &  48.73$\pm$16.70 &  35.62$\pm$  0.94&  2.02$\pm$ 0.32 &  4.14$\pm$ 0.75 & 21.5$\pm$ 1.0   & 34.73$\pm$ 0.90  &   4.97$\pm$  1.13  & 0 \\ 
 68 &  15.61$\pm$4.90 &  13.21$\pm$  1.47&  1.59$\pm$ 0.50 &  3.98$\pm$ 0.70 & 18.5$\pm$ 0.6   &  4.16$\pm$ 0.04  &   1.09$\pm$  0.12  & 0 \\ 
 78 &  34.02$\pm$10.45 &  34.25$\pm$  4.91&  0.46$\pm$ 0.06 &  0.90$\pm$ 0.12 & 17.6$\pm$ 1.1   & 33.03$\pm$ 0.89  &  39.27$\pm$  8.90  & I \\ 
 83 &  12.95$\pm$3.85 &  13.50$\pm$  1.84&  0.94$\pm$ 0.18 &  2.13$\pm$ 0.38 & 16.4$\pm$ 1.7   &  4.13$\pm$ 0.07  &   1.22$\pm$  0.23  & 0 \\ 
 85 &	12.65$\pm$3.90 &  12.82$\pm$  1.73&  0.88$\pm$ 0.18 &  2.42$\pm$ 0.38 & 16.2$\pm$ 1.6   &  4.19$\pm$ 0.11  &   1.04$\pm$  0.30  & 0 \\ 
 86 &	2.48$\pm$0.75 &   2.72$\pm$  0.32&  0.58$\pm$ 0.22 &  2.48$\pm$ 0.52 & 10.2$\pm$ 0.5   &  4.31$\pm$ 0.08  &   1.37$\pm$  0.24  & 0 \\ 
 87 &  22.14$\pm$8.80 &  16.83$\pm$  0.54&  2.74$\pm$ 0.43 &  6.15$\pm$ 0.86 & 19.6$\pm$ 0.1   &  4.32$\pm$ 0.18  &   2.88$\pm$  0.90  & 0 \\ 
 91 &  4.91$\pm$1.55 &   4.77$\pm$  0.70&  0.59$\pm$ 0.20 &  1.12$\pm$ 0.25 & 16.0$\pm$ 0.9   &  4.21$\pm$ 0.07  &   1.12$\pm$  0.33  & 0 \\ 
 92 &  6.60$\pm$2.00 &   6.10$\pm$  0.86&  1.10$\pm$ 0.25 &  2.13$\pm$ 0.31 & 13.1$\pm$ 0.4   &  4.15$\pm$ 0.03  &   0.94$\pm$  0.23  & 0 \\ 
98 &  3.82$\pm$1.05 &   3.88$\pm$  0.51&  0.41$\pm$ 0.16 &  0.96$\pm$ 0.23 & 11.0$\pm$ 0.7   &  4.19$\pm$ 0.06  &   1.10$\pm$  0.27  & 0 \\ 
99 &	4.59$\pm$1.45 &   4.64$\pm$  0.65&  0.55$\pm$ 0.18 &  1.06$\pm$ 0.24 & 12.0$\pm$ 1.0   &  4.19$\pm$ 0.07  &   1.10$\pm$  0.20  & 0 \\ 
102 &  1.91$\pm$0.65 &   2.02$\pm$  0.23&  0.12$\pm$ 0.06 &  1.06$\pm$ 0.19 &  9.3$\pm$ 0.5   &  4.32$\pm$ 0.08  &   2.76$\pm$  1.09  & 0 \\ 
104 &  1.69$\pm$0.60 &   1.69$\pm$  0.26&  0.12$\pm$ 0.02 &  0.45$\pm$ 0.18 &  8.5$\pm$ 0.6   &  4.22$\pm$ 0.08  &   1.31$\pm$  0.89  & 0 \\ 
105 &	3.07$\pm$0.95 &   2.64$\pm$  0.27&  0.02$\pm$ 0.00 &  0.26$\pm$ 0.02 &  7.8$\pm$ 0.2   & 22.20$\pm$ 0.29  &  13.60$\pm$  2.74  & I \\ 
106 &  3.78$\pm$1.10 &   3.58$\pm$  0.51&  0.01$\pm$ 0.00 &  0.21$\pm$ 0.01 &  8.4$\pm$ 0.3   & 22.83$\pm$ 0.67  & 131.77$\pm$ 23.42  & I \\ 
\enddata
\tablenotetext{a}{The L$_{bol}^0$ is calculated at 7 Kpc, while its range limits are calculated for 7$\pm$1~Kpc.}
\end{deluxetable}

\begin{table*}
\begin{center}
\caption{Herschel counterparts for the B10 and NS03 sources}
\begin{tabular}{cccccc}
\hline
Source ID & R.A. & Dec & Her-ID &  offset\\
\hline
 & [hh:mm:ss] & [dd:mm:ss] &  & arcsec  \\
\hline
  B-2 &  11:15:05.67 & -61:09:41.00 &  60 & 3.3 \\
  B-4 &  11:15:02.58 & -61:15:48.80 &  49 & 4.3\\
  B-5 &  11:15:10.18 & -61:16:12.70 &  78  & 14.0 \\
  B-6 &  11:15:08.88 & -61:16:54.80 &  72  &  3.5 \\
\hline 
  NS-1  &  11:15:04.67 & -61:09:37.6 & 59   & 3.2\\
  NS-2  & 11:15:05.88  & -61.09:40.0 & 60   &  1.5\\
  NS-4  & 11:15:03.79  & -61:14:22.4 & 55-56   &  3.8-4.1\\
  NS-9A & 11:15:11.34 & -61:16:45.2 & 81    & 9.8\\
  NS-12 & 11:15:6.50 & -61:17:56.8 & 64-65    & 6.3-5.8\\
  NS-13A & 11:15:11.34 & -61:20:25.1 & 87  & 7.0\\
  NS-13B & 11:15:11.22 & -61:20:36.7 & 87  & 6.2\\
  NS-15 & 11:15:07.00 & -61:10:08.9 & 68 & 5.8\\
  NS-28 & 11:15:16.10 & -61:16:05.4 & 95   & 7.16\\
  NS-36 & 11:15:14.81 & -61:17:00.0 & 93  & 6.0\\
  NS-48 & 11:15:08.94 & -61:21:01.7 & 77   & 7.3\\
\label{maser}
\end{tabular}
\end{center}
\end{table*}


\begin{figure*}
\centering
\includegraphics[width=1.\columnwidth]{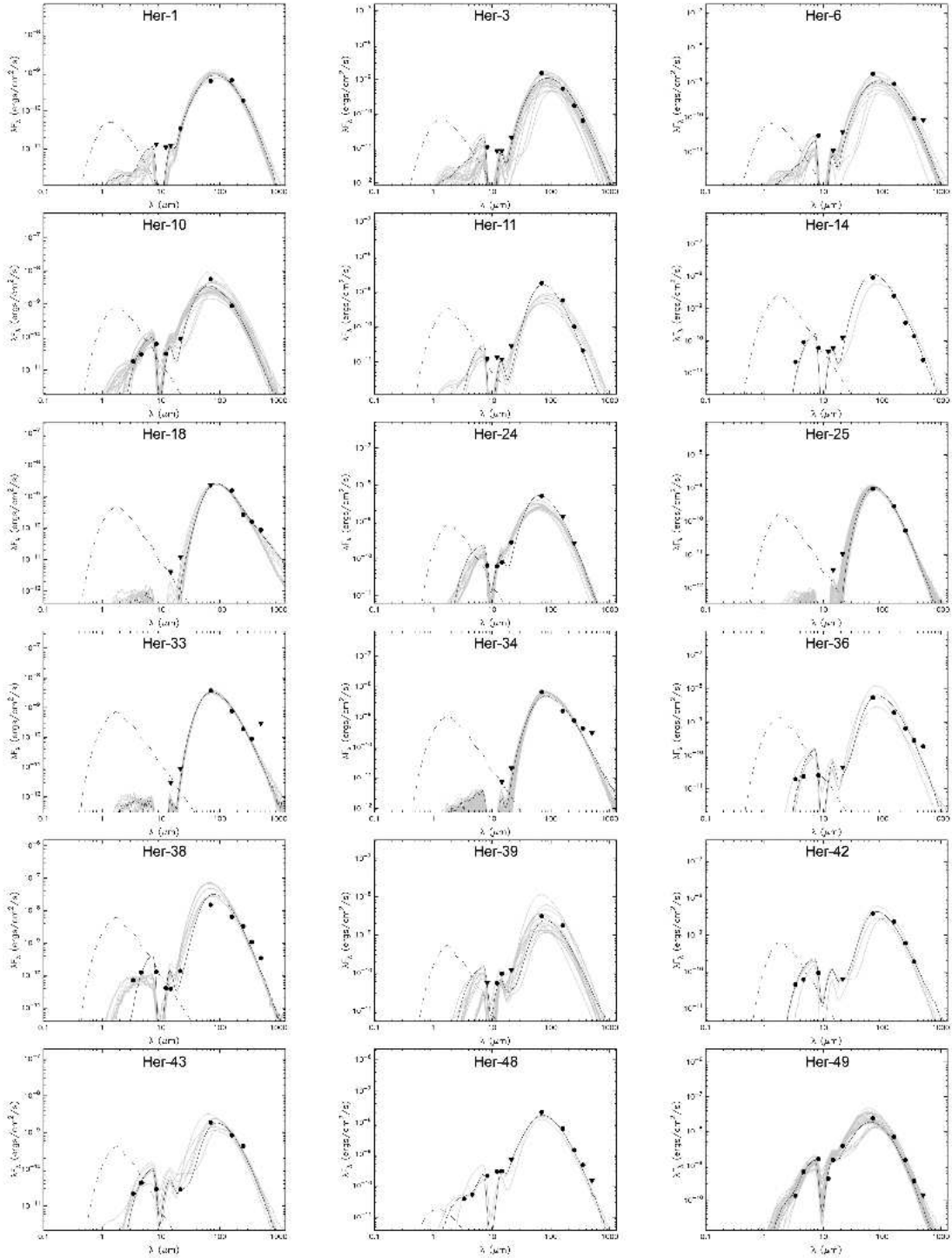}
\caption{SED models for the sources listed in Tab.\ref{sed}. The solid black line indicates the best model, and the gray lines show all the selected models. Black dashed line shows the SED of the stellar photosphere of the best fitting model (see R07).}
\label{f7}
\end{figure*} 

\setcounter{figure}{6}
\begin{figure}
\centering
\includegraphics[width=1.\columnwidth]{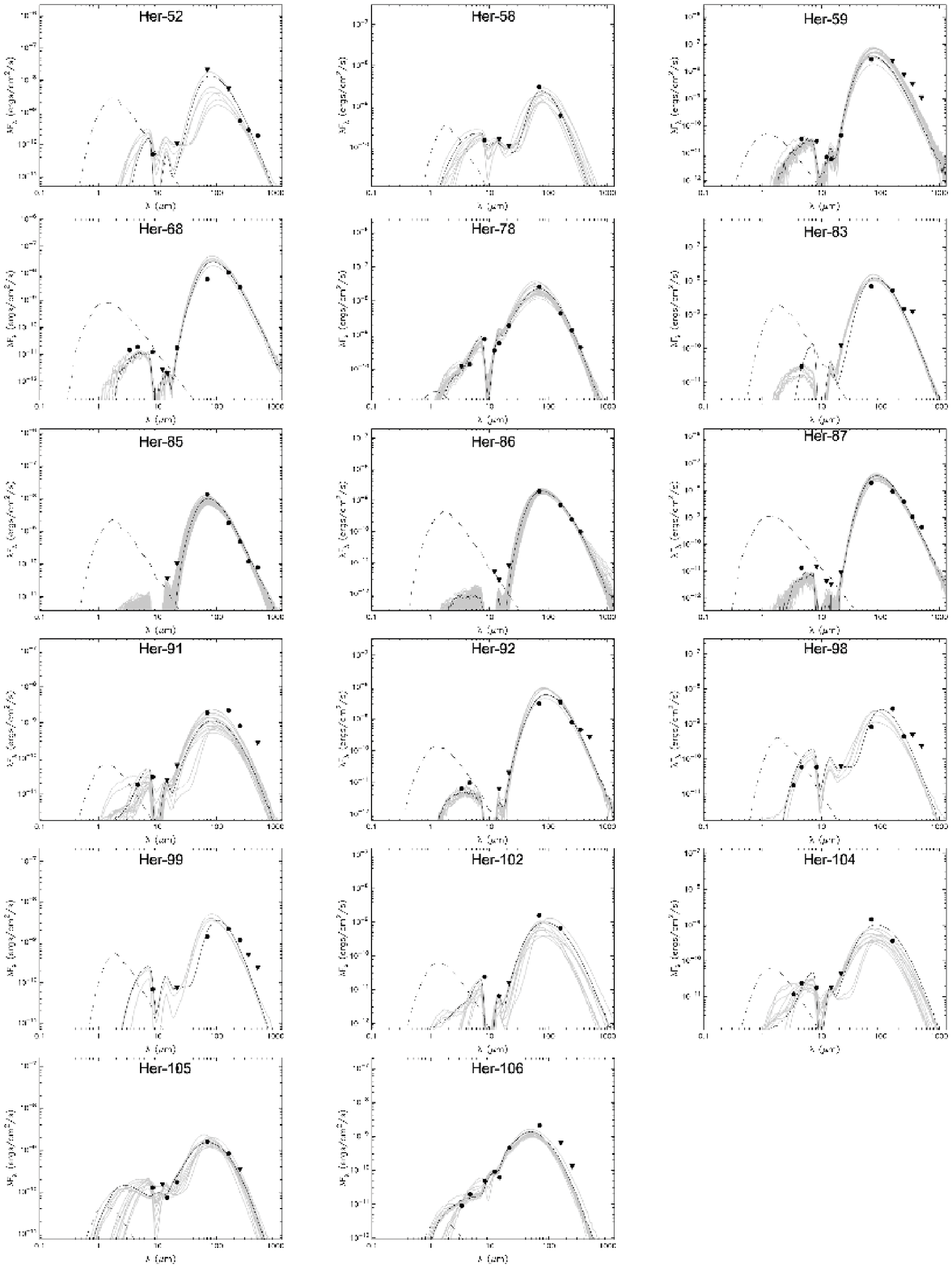}
\caption{continued}
\label{f8}
\end{figure}

\begin{figure}
\includegraphics[scale=0.6, angle=90]{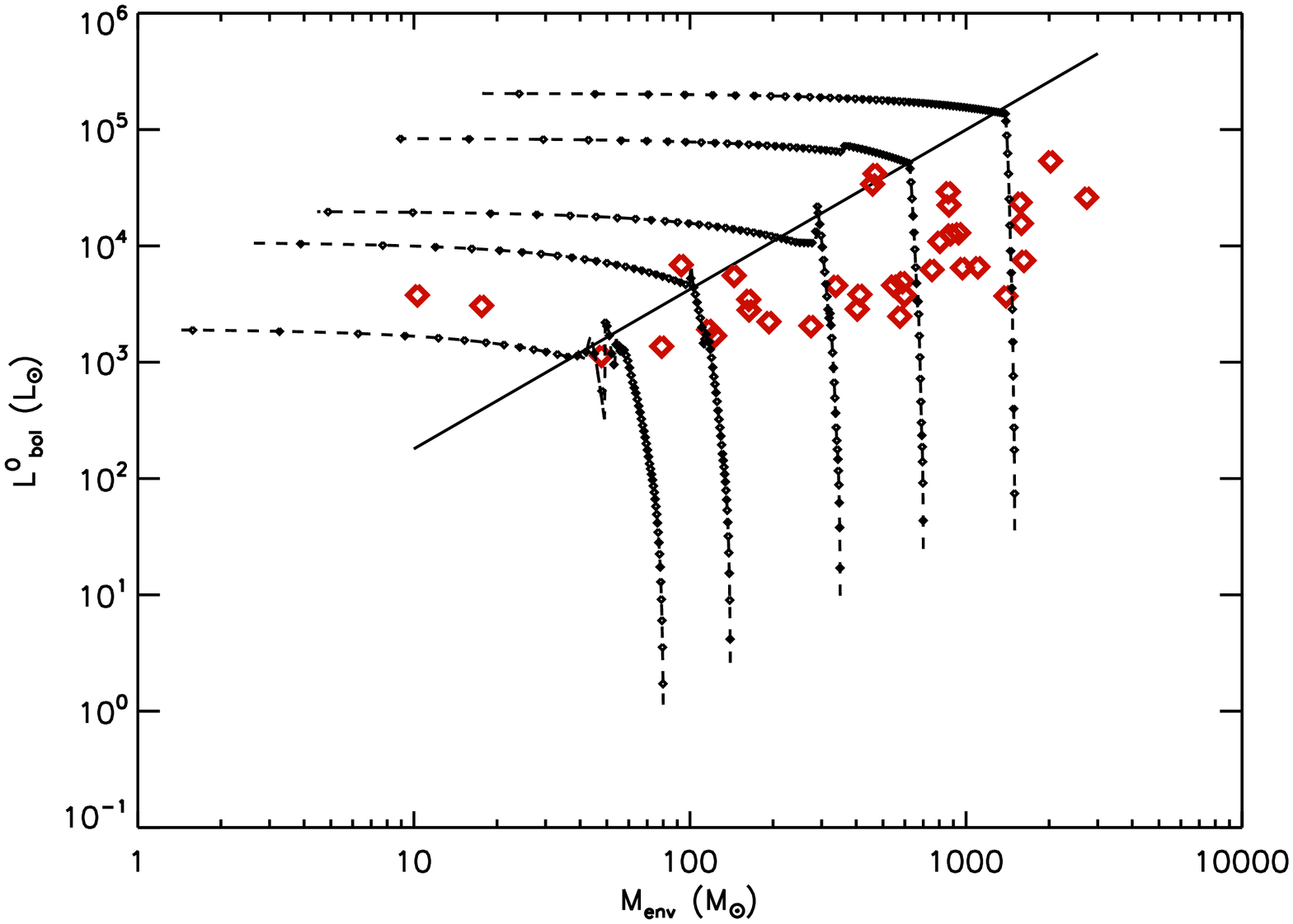}
\caption{$L_{bol}^O-M_{env}$ diagram for the sources (red squares) listed in Tab.\ref{sed}.  The black dashed lines mark the theoretical tracks calculated by \citet{mol08}, which are distinguished by the initial envelope mass ($M_{env}/M_{\sun}= 80, 140, 350, 700, 1500$). The diagonal black line highlights the moment when the YSOs arrive on the ZAMS, dividing the diagram in the earliest phases (vertical part of the tracks) and in the more evolved phases (horizontal part of the tracks).} 
\label{f9}
\end{figure}
 
\acknowledgments
{\it Acknowledgements}.  We thank the anonymous referee for his/her useful comments and suggestions to improve the results of this investigation. We are very grateful to Yasuo Fukui and his collaborators for kindly providing us the fits images related to their investigation (F14). We also thank the Hi-Gal Consortium team and in particular Sergio Molinari for the long time collaboration and for providing us the calibration maps.

The Herschel data used in this paper belong to the fields Field290\_ 0 (Obs-ID: 1342203081, 1342203082) and Field292\_ 0 (Obs-ID: 1342203064, 1342203065), acquired during the Observing Days 458 and 459 respectively. Herschel data are available at the following webpage:  http://www.asdc.asi.it~/mmia/index.php?~mission=herschel. 

The Herschel spacecraft was designed, built, tested, and launched under a contract to ESA managed by the Herschel/Planck Project team by an industrial consortium under the overall responsibility of the prime contractor Thales Alenia Space (Cannes), and including Astrium (Friedrichshafen) responsible for the payload module and for system testing at spacecraft level, Thales Alenia Space (Turin) responsible for the service module, and Astrium (Toulouse) responsible for the telescope, with in excess of a hundred subcontractors.

This research made use of data products from the Midcourse Space Experiment. Processing of the data was funded by the Ballistic Missile Defense Organization with additional support from NASA Office of Space Science. 
This research has also made use of the NASA/IPAC Infrared Science Archive, which is operated by the Jet Propulsion Laboratory, California Institute of Technology, under contract with the National Aeronautics and Space Administration.

This publication makes use of data products from the Wide-field Infrared Survey Explorer, which is a joint project of the University of California, Los Angeles, and the Jet Propulsion Laboratory/California Institute of Technology, funded by the National Aeronautics and Space Administration.
 
\vspace*{\fill}
\newpage

\bibliographystyle{apj}

\begin{thebibliography}{100}

\bibitem[Aniano et al.(2011)]{aniano11}
Aniano, G., Draine, B.T., Gordon, K.D., \& Sandstrom, K., 2011, PASP, 123, 1218

\bibitem[Beccari et al.(2010)]{beccari10} Beccari, G., Spezzi, L., De Marchi, G. et al., 2010, ApJ, 720, 1108 
 

\bibitem[Bohlin, Savage \& Drake(1978)]{bohlin78} Bohlin, R. C., Savage, B. D., \& Drake, J. F. 1978, ApJ, 224, 132 

\bibitem[Brandl et al.(1999)]{brandl99} Brandl, B., Brandner, W., Eisenhauer et al. 1999, A\&A, 352, L69

\bibitem[Brandner et al.(2000)]{brandner00} Brandner, W., Grebel, E. K., Chu, Y.-H., et al. 2000, AJ, 489, 292

\bibitem[Breen et al.(2010)]{breen10}
Breen, S.L., Caswell, J.L., Ellingsen, S.P., \& Phillips, C.J., 2010, MNRAS,
406, 1487 (B10)

\bibitem[Caswell et al.(1989)]{caswell89} Caswell, J. L., Batchelor, R.A., Forster, J. R. et al. 1989, Australian Journal of Physics, 42, 331

\bibitem[Caswell(2004)]{caswell04}
Caswell, J.L., 2004, MNRAS, 351, 279

\bibitem[Correnti et al.(2012)]{correnti12}
Correnti, M.,  Paresce, F., Aversa, R. et al., 2012, Ap\&SS, 340, 263 

 \bibitem[de Pree, Nysewander \& Goss(1999)]{depree99} de Pree, C. G., Nysewander, M. C. \& Goss, W. M. 1999, AJ, 117, 2902

\bibitem[Draine (1978)]{draine78} Draine, B. T. 1978, ApJS, 36, 595

\bibitem[Draine \& Lee(1984)]{draine84}Draine, B. T., \& Lee, H. M. 1984, ApJ, 285, 89 

\bibitem[Draine \& Li(2007)]{draine07}
Draine, B.T., \& Li, A., 2007, ApJ, 657, 810 (DL07)

\bibitem[Egan, Price \& Kraemer(2003)]{egan03}Egan, M. P., Price, S. D., \& Kraemer, K. E. 2003, AAS, 203, 5708

\bibitem[Etxaluze et al. (2013)]{etxaluze} Etxaluze, M., Goicoechea, J. R., Cernicharo, J. et al. 2013, A\&A, 556, 137

 \bibitem[Frogel, Persson \& Aaronson (1977)]{frogel}Frogel, J. A., Persson, S. E.,\& Aaronson, M.
  1977, ApJ, 213, 723

\bibitem[Fukui et al. (2014)]{fukui}Fukui, Y., Ohama, A., Hanaoka, N. et al. 2014, ApJ, 780, 36 (F14)
 
\bibitem[Gaczkowski et al(2013)]{gat13} Gaczkowski, B., Preibisch, T., Ratzka, T., et al. 2013 A\&A, 549, 47


\bibitem[Georgelin et al. (2000)]{georgelin}Georgelin, Y. M., Russeil, D., Amram, P. et al.  2000, A\&A, 357, 308


\bibitem[Grabelsky et al.(1988)]{grab88}	Grabelsky, D. A., Cohen, R. S., Bronfman, L., \& Thaddeus, P. 1988, ApJ, 331, 181


\bibitem[Griffin et al.(2010)]{griffin10}
Griffin, M.J., Abergel, A., Abreu, A. et al., 2010, A\&A, 518, L3

\bibitem[Harayama et al.(2008)]{harayama08} 
Harayama, Y., Eisenhauer, F., \& Martins, F., 2008, ApJ, 675, 1319  


 \bibitem[Lebouteiller et al. (2007)]{lebo07} Lebouteiller, V., Brandl,. B., Bernard-Salas, J. et al. 2007, ApJ, L665, 390	(L07)
 
 \bibitem[Lebouteiller et al. (2008)]{lebo08} Lebouteiller, V., Bernard-Salas, J., Brandl, B. et al. 2008, ApJ, L680, 398

\bibitem[Melena et al.(2008)]{melena08} 
Melena, N.W., Massey, P., Morrell, N.I., \& Zangari, A.M., 2008, AJ, 135, 878

\bibitem[Melnick, Tapia \& Terlevich(1989)]{melnick89} 
Melnick, J., Tapia, M., \& Terlevich, R., 1989, A\&A, 213, 89 

\bibitem[Moffat (1983)]{moffat83} Moffat, A. F. J., Abergel, A.,  Abreu, A. et al. 1983, A\&A, 124, 273 

\bibitem[Moffat et al.(2002)]{moffat02}
Moffat, A.F.J., et al., 2002, ApJ, 573, 191

\bibitem[Molinari et al. (2008)]{mol08} Molinari, S., Pezzuto, S., Cesaroni, R. et al. 2008,  A\& A, 481, 345

\bibitem[Molinari et al.(2010)]{mol10} Molinari, S., Swinyard, B., Bally, J., et al., 2010, A\& A, 518, 100 (MO10)


\bibitem[Molinari et al.(2010b)]{mol10b} 
Molinari, S., Swinyard, B., Bally, J. et al.  2010b, PASP, 122, 314 

\bibitem[Molinari et al.(2011)]{mol11}
Molinari, S., Schisano, E., Faustini, F., et al., 2011, A\& A,530,133 (MO11)

\bibitem[N\"urnberger et al.(2002)]{nurn02b} 
N\"urnberger, D.E.A., Bronfman, L., Yorke, H.W., \& Zinnecker, H., 2002, A\&A, 
394, 253 (N02)

\bibitem[N\"urnberger \& Petr-Gotzens(2002)]{nurn02} 
N\"urnberger, D.E.A., \& Petr-Gotzens, M.G., 2002, A\&A, 382, 537 

\bibitem[N\"urnberger (2003)]{nurn03b}   N\"urnberger, D.E.A., 2003, A\&A, 404, 255 

\bibitem[N\"urnberger \& Stanke(2003)]{nurn03}N\"urnberger, D.E.A., \& Stanke, T., 2003, A\&A, 400, 223 (NS03)

\bibitem[N\"urnberger (2008)]{nurn08} N\"urnberger, D.E.A., 2008, JPhCS, 131, A2025  

\bibitem[Oldam et al.(1994)]{oldham94} Oldham, P. G., Griffin, M. J., Richardson, K. J., \& Sandell, G. 1994, A\& A, 284, 559O

\bibitem[Ossenkopf \& Henning(1994)]{oh94} Ossenkopf, V., \& Henning, T. 1994, A\&A, 291,943 (OH94)

\bibitem[Pang, Pasquali \& Grebel(2011)]{xiao11} Pang, X, Pasquali, \& A, Grebel  2011, AJ, 142, 132 (P11) 

\bibitem[Persi et al.(1994)]{persi94} Persi, P., Roth, M., Tapia, M. et al. 1994, A\&A, 282, 474

\bibitem[Piazzo et al.(2012)]{piaz12}
Piazzo, L., et al., 2012, IEEE Trans. Image Proc., 21, 8, 3687

\bibitem[Pilbratt et al.(2010)]{pilbratt10}	Pilbratt, G. L., Riedinger, J. R., Passvogel, T.
Pilbratt, G.L., et al. 2010, A\&A, 518, L1

\bibitem[Poglitsch et al.(2010)]{pog10} Poglitsch, A., Waelkens, C., Geis, N. et al. 2010, A\&A, 518, L2 

\bibitem[Preibisch et al.(2012)]{preibisch12} 
Preibisch, T., Roccatagliata, V., Gaczkowski, B., \& Ratzka, T., 2012, A\& A,
541, 132 

\bibitem[Robitaille et al.(2006)]{rob06} Robitaille, T. P., Whitney, B. A., Indebetouw, R. et al. 2006, ApJS, 167, 256

\bibitem[Robitaille et al (2007)]{rob07} Robitaille, T. P., Whitney, B. A., Indebetouw, R., \& Wood, K. 2007, ApJS, 169, 328 (RB07)

\bibitem[Rochau et al.(2010)]{rochau} Rochau, B., Brandner, W., Stolte, A. et al. 2010, ApJ, 716, L90

\bibitem[R\"olling et al.(2011)]{roll11} R\"ollig, M., Kramer, C., Rajbahak, C. et al. 2011, A\& A, 525, 8 (R11)



\bibitem[Scheen et al.(2006)]{Scheen06} 
Schnee, S., Bethell, T., \& Goodman, A. 2006, ApJ, L640, 47
 
\bibitem[Tapia et al.(2001)]{tapia01} Tapia, M., Bohigas, J., P\'erez, B. et al. 2001, RMxAA, 37, 39 

\bibitem[Townsley et al.(2011)]{town11} Townsley, L. K. Broos, P. S., Chu, You-Hua et al.  2011, ApJS, 194, 16
	
\bibitem[Traficante et al.(2011)]{traf11} 
Traficante, A., Calzoletti, L., Veneziani, M. et al., 2011, MNRAS, 416, 2932

\bibitem[Urquhart, Morgan \& Thompson(2009)]{urq09} Urquhart, J. S., Morgan, L. K., \& Thompson, M. A. 2009, A\&A, 497, 789

\bibitem[Vehoff et al. (2010)]{Veo10} Vehoff, S., Hummel, C.A., Monnier, J.D., et al., 2010, A\&A, 520, A78 

\bibitem[Wang \& Chen(2010)]{WangChen10} Wang, J., \& Chen, Y. 2010, Science China Phys., 53S, 271

\bibitem[Ward-Thompson, Andr\'e \& Kirk(2002)]{ward02}Ward-Thompson, D., Andr\'e, P.,\& Kirk, J. M. 2002, MNRAS, 329, 257

\bibitem[Whitney et al.(2003)]{whi03} Whitney, B. A., Wood, K., Bjorkman, J. E., \& Wolff, M. J. 2003, ApJ, 591, 1049

\bibitem[Whitney et al.(2003b)]{whi03b} Whitney, B. A., Wood, K., Bjorkman, J. E., \& Cohen, M. 2003, ApJ, 598, 1079

\bibitem[Wright et al.(2010)]{wise}Wright, E. L., Eisenhardt, P. R. M, Mainzer, A. K. 2010, AJ, 140, 1868


\end{thebibliography}

\end{document}